\documentclass[aps,prl,superscriptaddress,reprint]{revtex4-1}
\usepackage{amsmath}
\usepackage{amssymb}
\usepackage{graphicx}
\usepackage[colorlinks,urlcolor=blue]{hyperref}
\usepackage{url}
\usepackage{color,xcolor}
\usepackage{ulem}
\usepackage{float}
\usepackage{epstopdf}
\usepackage[none]{hyphenat}
\begin{document}

\title{Superconductivity in Ruddlesden-Popper nickelates: a review of recent progress, focusing on thin films}
\author{Yang Zhang}
\affiliation{Materials Science and Technology Division, Oak Ridge National Laboratory, Oak Ridge, Tennessee 37831, USA}
\email{zhangy10@ornl.gov}
\author{Ling-Fang Lin}
\affiliation{Department of Physics and Astronomy, University of Tennessee, Knoxville, Tennessee 37996, USA}
\author{Thomas A. Maier}
\affiliation{Computational Sciences and Engineering Division, Oak Ridge National Laboratory, Oak Ridge, Tennessee 37831, USA}
\author{Elbio Dagotto}
\email{edagotto@utk.edu}
\affiliation{Department of Physics and Astronomy, University of Tennessee, Knoxville, Tennessee 37996, USA}
\affiliation{Materials Science and Technology Division, Oak Ridge National Laboratory, Oak Ridge, Tennessee 37831, USA}
\date{\today}

\begin{abstract}
The discovery of superconductivity with $T_c \sim 80$ K in the nickelate Ruddlesden-Popper bilayer La$_3$Ni$_2$O$_7$ at high pressure has opened a new platform for unconventional superconductivity, followed by the subsequent observation of superconductivity in trilayer La$_4$Ni$_3$O$_{10}$, also at high pressure. Remarkably, ambient-pressure superconductivity was also observed recently in La$_3$Ni$_2$O$_7$ ultra-thin films when grown on substrates that provide compressive strain. This discovery significantly extends the type of experimental techniques that can be used in nickelates, previously limited due to the high-pressure constraint. Discussing the similarities and differences among these nickel oxides will provide new insights into understanding the mechanism of high-$T_c$ superconductivity in correlated electron systems. In this paper, we review the experimental and theoretical progress on Ruddlesden–Popper nickelates, with emphasis on thin films, and discuss future perspectives and research directions.
\end{abstract}

\maketitle

\section{Introduction}
\label{sec_1}
\vspace{6pt}

Since the initial discovery of superconductivity in cuprates in the mid-late 1980s~\cite{Bednorz:Cu}, unconventional high critical temperature (``high $T_c$'') superconductivity in strongly correlated electron systems has attracted considerable sustained attention and remains a central topic in condensed matter physics. As the CuO$_2$ layers are the crucial ingredients for high $T_c$ superconductivity in those Cu-based oxides~\cite{Dagotto:rmp94,Scalapino:rp95}, the search for other layered oxides exhibiting electronic structures similar to those of the cuprates has been a long-standing strategy and formidable challenge.

After decades of studies, a new transition metal nickel oxide has recently emerged in the challenging field of high $T_c$ unconventional superconductivity after the extremely successful discovery of high $T_c$ in the Cu oxides. In between these developments, high $T_c$ was also reported in iron-based materials, though primarily in pnictides and chalcogenides rather than oxides~\cite{Kamihara:jacs,Dai:np12,Dagotto:Rmp,Dai:Rmp15,Dagotto:extra1}. Thus, currently there are three families of compounds where robust $T_c$ has been observed, well above the limit expected for conventional phonon-mediated Bardeen–Cooper–Schrieffer (BCS) superconductors: the Cu-, Fe-, and Ni-based superconductors. Current efforts in the community focus on clarifying whether (1) the pairing mechanism is indeed magnetic, as widely assumed, and (2) which pairing channel gives rise to superconductivity.

In this review, we focus on recent advances in the emerging field of Ruddlesden-Popper (RP) nickelate superconductors, with particular emphasis on {\it ultra-thin film systems}. The structure of the review is as follows.

First, we briefly review the groundbreaking discovery of superconductivity in the NdNiO$_2$ thin films in Sec.2. Second, in Sec.3, we discuss the exciting discoveries and growing interest in bilayer and trilayer nickelates since 2023, following reports of superconductivity at approximately $\sim 80$ K under high pressure in bulk crystals. In Sec.4, turning to the main focus of this review article, we will address the key experimental results in the subfield of ultra-thin films within the RP bilayer family of compounds. This subfield started in 2025 when superconductivity in Ni-oxide films of the La$_3$Ni$_2$O$_7$ material was reported. The major advance in this context was the observation of superconductivity in ambient-pressure samples, enabling a broader range of experimental investigations that were previously impossible under high-pressure conditions, such as those involving angle-resolved photoemission (ARPES) techniques in the superconducting state. Sec.5 of the review presents theoretical developments in the ultra-thin film context. Then, we also discuss recent progress on hybrid stacking RP nickelates in Sec.6. Finally, Sec.7 provides an overview, summarizes the main conclusions, and outlines directions for future research.  For other reviews on RP nickelates, see references~\cite{Wang:cplreview,Goodge:pt2025,Wang:nsr2025,Puphal:nrp2025,Yao:extra,Oh:extra}.

\section{Synthesis of superconducting NdNiO$_2$ thin films}
\label{sec_2}
\vspace{6pt}

\begin{figure*}
\centering
\includegraphics[width=0.88\textwidth]{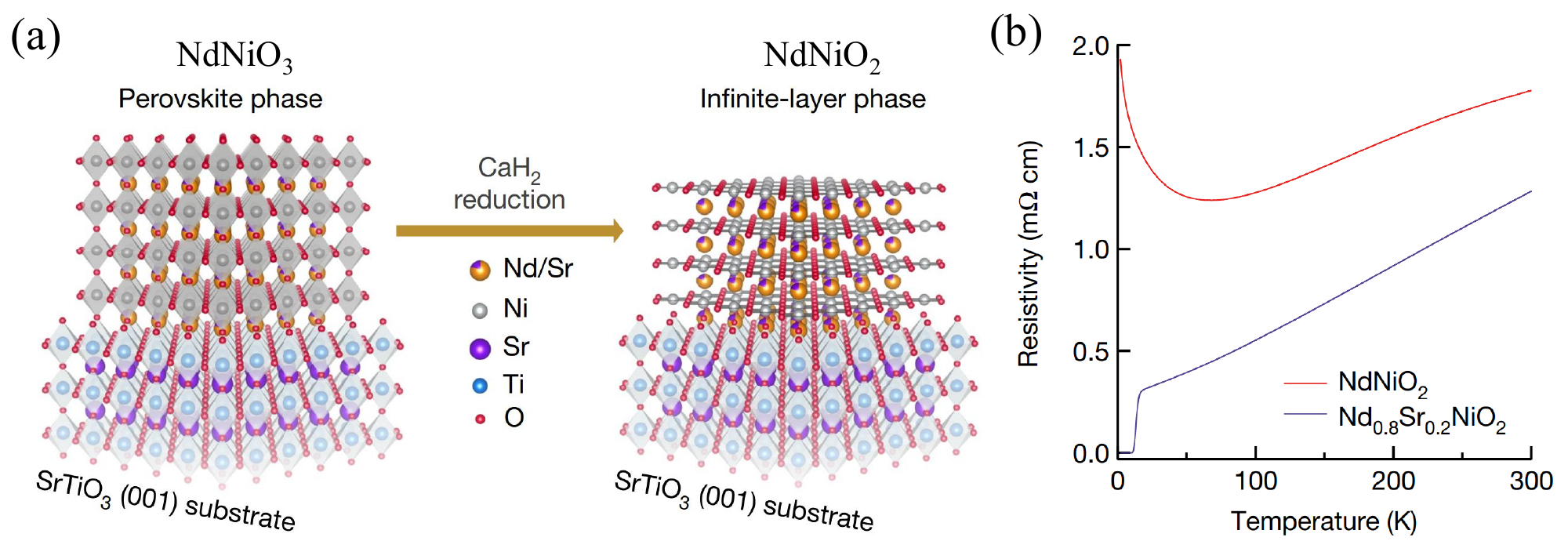}
\includegraphics[width=0.88\textwidth]{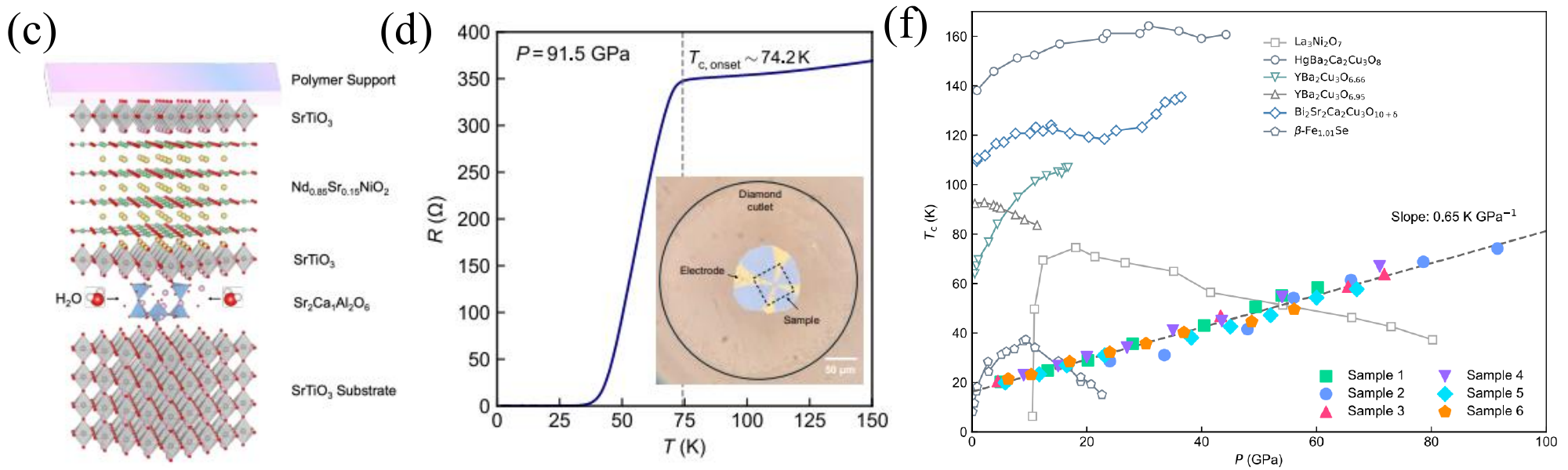}
\caption{ (a) Sketch of the oxygen reduction process used to convert NdNiO$_3$ thin films into NdNiO$_2$ thin films, where in the latter Ni is in a rare 1+ valence state. In this example, the substrate is STO, though other substrates are discussed later in this review. (b) Temperature-dependent resistivity of both NdNiO$_2$ and hole-doped Nd$_{0.8}$Sr$_{0.2}$NiO$_2$ thin-film samples. The latter shows superconductivity at $\sim 15$ K. (a-b) are reprinted with permission from Springer Nature Customer Service Centre GmbH: Springer Nature, Nature, ~\cite{Li:Nature}, Copyright (2019). (c) Schematic of Nd$_{0.85}$Sr$_{0.15}$NiO$_2$ thin films ($\sim 6.7$ nm) encapsulated by protective STO capping layers and grown on a water-soluble sacrificial layer. (d) Temperature-dependent resistivity $R$ vs. $T$ for the film in (c) at $P = 91.5$ GPa, showing superconductivity around $\sim 74.2$ K. Inset: an optical microscope image of the sample chamber containing the sample. (f) Superconducting critical temperature vs. pressure for the thin films in (c), compared with other families of high-$T_c$ superconductors. (c-f) are reproduced from~\cite{Lee:arxiv2026}, CC BY 4.0.}
\label{fig1}
\end{figure*}

The recent surge of interest in Ni-oxide high-$T_c$ superconductivity was sparked by a landmark study in 2019~\cite{Li:Nature}, where superconductivity was reported with a critical temperature of $T_c \sim 15$ K in NdNiO$_2$ thin films after hole doping through partial substitution of Nd$^{3+}$ with Sr$^{2+}$. As shown in Fig.~\ref{fig1}(a), the preparation of the thin films involves a multi-step process. The first step is the growth of NdNiO$_3$ thin films on commonly used oxide substrates, such as  SrTiO$_3$ (STO), employing canonical layer-by-layer growth techniques, including pulsed laser deposition or molecular beam epitaxy. In the NdNiO$_3$ perovskite phase, each Ni ion is coordinated by six oxygen atoms in a characteristic octahedral geometry. NdNiO$_3$ itself is not superconducting; instead, it is an antiferromagnetic insulator with spin canting~\cite{Kumar:prb13}.

As illustrated in Fig.~\ref{fig1}(a), after undergoing an oxygen reduction process, where a fraction of the oxygen is removed chemically (reduction), the original NdNiO$_3$ thin film is transformed into NdNiO$_2$. Of the six oxygen ions originally coordinating each Ni ion, only four in-plane oxygens remain after the reduction process. As a result, the perovskite structure transforms into the so-called infinite-layer phase, which cannot be directly grown as a single crystal due to structural instabilities. Therefore, an oxygen-reduction process is necessary to obtain this phase in thin-film form. Remarkably, this oxygen-reduced ``112'' compound becomes superconducting upon approximately $20 \%$ hole doping, by partial replacement of Nd$^{3+}$ with Sr$^{2+}$. This process transforms a weakly insulating NdNiO$_2$ into a Nd$_{0.8}$Sr$_{0.2}$NiO$_2$ superconductor at low temperature, as evidenced by the resistivity–temperature curves shown in Fig.~\ref{fig1}(b).

Although Ni$^{1+}$ ($d^9$) is isoelectronic with Cu$^{2+}$ ($d^9$), many studies have revealed fundamental differences between individual infinite-layer nickelates and cuprates (see e.g., Ref.~\cite{Li:prl20,Nomura:rpp,Zhang:prb20,Gu:innovation,Zhang:prb2025-il,Botana:prx,Wu:prb20,Karp:prx,Gao:nc2024-IL,Ren:cp2023-IL,Ren:cpl2024-IL}). This ``112'' nickelate belongs to the simplest $m = \infty$ system of the reduced RP perovskite $R_{m+1}$Ni$_m$O$_{2m+2}$ family. Interestingly, the high-order layer stacking ($m$ = 4-7) reduced RP nickelates also show superconductivity, such as Nd$_6$Ni$_5$O$_{12}$ ($d^{8.8}$) with $T_c \sim 13$ K~\cite{Pan:nm}. The whole superconducting phase diagram of multi-layer structures has recently been reported~\cite{Pan:science}. Here, we do not attempt to review the extensive body of work on these ``112'' and other reduced RP nickelates, but for proper context, note that, until recently, the record critical temperature of ambient-pressure superconductivity in the ``112'' family is approximately 40 K, recently achieved in hole-doped SmNiO$_2$~\cite{Chow:Nature}. Notably, Ref.~\cite{Chow:Nature} reports resistivity vs temperature curves exhibiting nearly linear behavior for the ``112'' compound studied there, resembling results discussed later in this review for ultra-thin films of bilayer nickel oxides.

Very recently, Nd$_{0.85}$Sr$_{0.15}$NiO$_2$ optimally-doped  thin films ($\sim 6.7$ nm) were synthesized via pulsed laser deposition and topotactic reduction~\cite{Lee:arxiv2026}. The film structure involves a polymer-supported STO (10 u.c.)/Nd$_{0.85}$Sr$_{0.15}$NiO$_2$ (20 u.c.)/STO(10 u.c.) /Sr$_2$CaAl$_2$O$_6$ heterostructure grown on the STO substrate, followed by water etching of the sacrificial layer, as displayed in Fig.~\ref{fig1}(c). By increasing the pressure, as displayed in Fig.~\ref{fig1}(d), a record critical temperature of high-pressure superconductivity with $T_c \sim 74.2$ K~\cite{Lee:arxiv2026} was reported in these thin films at 91.5 GPa (the maximum pressure allowed by their equipment). Figure~\ref{fig1}(f) summarizes the pressure dependence $T_c(P)$ for several samples, comparing with other families of compounds, from Ref.~\cite{Lee:arxiv2026}. Interestingly, a simple linear enhancement of $T_c$ with a slope of 0.65 K GPa$^{-1}$, without signs of saturation, was observed.

\section{Bulk superconductivity in the new RP Nickel Oxides}
\label{sec_3}
\vspace{6pt}

\subsection{Superconductivity in bilayer La$_3$Ni$_2$O$_7$ bulk under pressure}
\label{subsec_3-1}
\vspace{4pt}

Before turning to our primary focus on thin films of La$_3$Ni$_2$O$_7$ at ambient pressure, we briefly highlight the recent exciting developments in {\it bilayer} RP perovskite nickelates at high pressure, which triggered the recent frenzy in the field of Ni-based high-$T_c$ superconductors.

In 2023, a remarkable breakthrough for nickel oxides was reported by the group of Prof. Meng Wang at Sun Yat-sen University, China, where they found that La$_3$Ni$_2$O$_7$~\cite{Sun:Nature23}, a bilayer RP perovskite nickel oxide, exhibits superconductivity above the boiling point of liquid nitrogen ($\sim 77$ K). The critical temperature $T_c$ reaches approximately 80 K at 14 GPa~\cite{Sun:Nature23}. In this compound, nickel has an average valence of +2.5, which differs significantly from the Ni$^{1+}$ state in infinite-layer nickelates~\cite{Hou:arxiv,Wang:arxiv9,Wang:nature,Dong:arxiv12}. This non-integer valence implies that the Ni lattice is intrinsically {\it self-doped} with holes.

\begin{figure}
\centering
\includegraphics[width=0.46\textwidth]{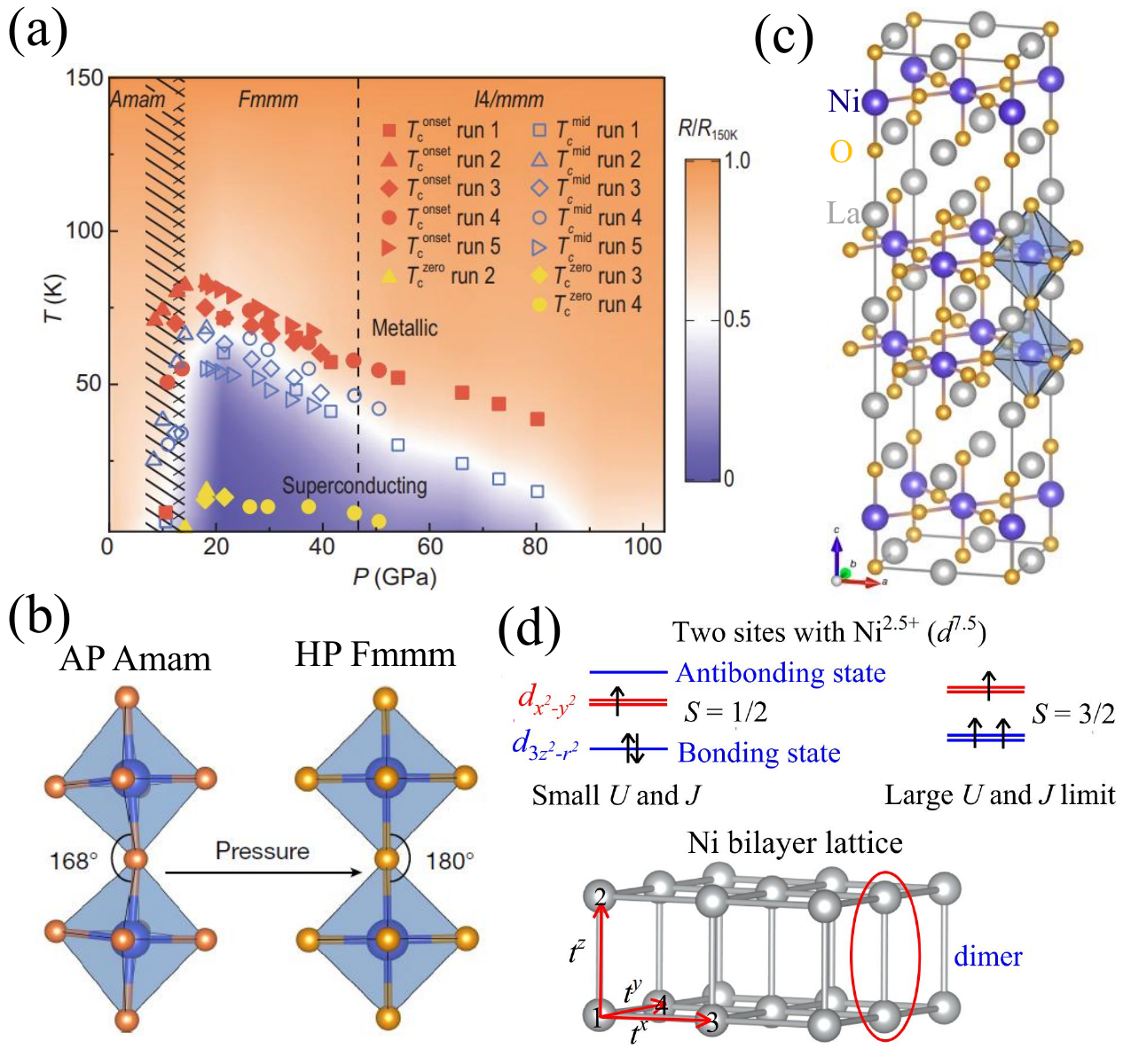}
\caption{ (a) Temperature–pressure phase diagram of La$_3$Ni$_2$O$_7$. Reprinted from~\cite{Li:NSR25}, National Science Review, 2025, CC BY 4.0. (b) Illustration of the two crystal structures of La$_3$Ni$_2$O$_7$: the low-pressure (LP) phase has $Amam$ symmetry, while the high-pressure (HP) phase has $Fmmm$ symmetry. (c) Crystal structure of the bilayer nickel oxide La$_3$Ni$_2$O$_7$, with Ni atoms shown in blue, oxygen in orange, and lanthanum in gray. Panels (b) and (c) are reprinted with permission from Springer Nature Customer Service Centre GmbH: Springer Nature, Nature~\cite{Sun:Nature23}, Copyright (2023).  (d) Top: the four electronic states of a two-Ni unit cell in the weak-coupling (small Hubbard $U$) and strong-coupling (large Hubbard $U$) regimes, considering only the $d_{x^2-y^2}$ and $d_{3z^2-r^2}$ orbitals.
Bottom: schematic of the bilayer geometry showing only Ni atoms; the ellipse highlights the location of the ``dimers'' discussed in the text. Reprinted with permission from~\cite{Zhang:prb23}, Copyright (2025) by the American Physical Society.}
\label{fig2}
\end{figure}

We now discuss the phase diagram of La$_3$Ni$_2$O$_7$ as a function of pressure. Here, hydrostatic pressure is applied, such that all three lattice parameters ($a$, $b$, and $c$) are reduced simultaneously. This should be contrasted with compressive strain, to be discussed later in the review, in which only the in-plane lattice constants ($a$ and $b$) are reduced, while the out-of-plane $c$ axis expands. As shown in Fig.~\ref{fig2}(a), the superconducting phase of La$_3$Ni$_2$O$_7$ extends over a wide pressure range, from about 14 to 90 GPa, with an optimal pressure near 20 GPa~\cite{Li:NSR25}. Recently, the experimental efforts have continued to improve the sample quality and, consequently, a transition temperature $T_c$ up to 100 K was obtained after partial replacement of La$^{3+}$ with other rare-earth 3+ elements, under pressure~\cite{Li:nature-96K,QiuZ:arxiv25}.

At ambient pressure (often associated with 0 GPa) and in the low-pressure region, superconductivity is absent; however, in this pressure range, anomalies in the resistivity vs. temperature curves indicate the formation of the spin-density wave (SDW) or charge-density wave~\cite{Sun:Nature23,Liu:scpma}. In particular, the SDW attracted considerable attention because it may be considered the parent compound of the superconducting state of La$_3$Ni$_2$O$_7$~\cite{Chen:arxiv2024,Chen:prl24,Dan:arxiv2024,Khasanov:NP2024}. The dominant magnetic features in the SDW phase will be briefly reviewed later in Sec.~\ref {subsec_3-4}. Because the low and high pressure regions are separated by a first-order phase transition, it is not clear that one state emerges from the other, as it occurs in cuprates when hole or electron doping of the parent antiferromagnetic compound, such as La$_2$CuO$_4$~\cite{Dagotto:rmp94}, leads to superconductivity. Further work is needed to clarify this matter.

Furthermore, a crucial difference between the low- and high-pressure regimes is the lattice distortion, as illustrated in Fig.~\ref{fig2}(b). At low pressure, the lattice adopts an orthorhombic $Amam$ structure with tilted octahedra, reducing to approximately 168$^\circ$ the Ni–O–Ni bond angle~\cite{Sun:Nature23}. At high pressure, the lattice becomes tetragonal, or nearly tetragonal, with untilted octahedra, restoring the Ni–O–Ni bond angle to 180$^\circ$ and adopting the $Fmmm$ structure. The transition between these phases corresponds to a first-order phase transition~\cite{Sun:Nature23}.

The conventional cell of La$_3$Ni$_2$O$_7$ is shown in Fig.~\ref{fig2}(c), where the bilayers are clearly visible, with each Ni atom coordinated by six oxygen atoms. Between the bilayers, there is a region populated by La and O ions, which can be regarded as a charge reservoir. Note that in between adjacent bilayers, there is a shift by 1/2 lattice spacings. Consequently, strictly speaking, two bilayers are needed per unit cell rather than just one, although the coupling between them is much smaller than the coupling between layers of the individual bilayers. This detail becomes particularly relevant when discussing thin films, as an ultra-thin film of one unit cell thickness will contain two bilayers, as will be addressed later in Sec.~\ref {sec_4}.

As widely discussed in the literature on bilayer nickelates~\cite{Luo:prl23,Zhang:prb23,Christiansson:arxiv,Zhang:prb23-2,Geisler:npjqm}, a proper description requires two Ni $3d$ orbitals: $d_{x^2-y^2}$ and $d_{3z^2-r^2}$. In a perovskite geometry, they correspond to the two highest-energy $3d$ orbitals. Although there is crystal-field energy in the octahedral environment, this energy gap is not large enough to allow for the neglect of the lowest energy $d_{3z^2-r^2}$ orbital. In addition, the Hund's coupling between these orbitals often plays a significant role. However, in general, it is widely believed that the three other $3d$ orbitals, namely $d_{xy}$, $d_{yz}$, and $d_{xz}$, are doubly occupied and, thus, not needed for modelling.

In addition, bilayer La$_3$Ni$_2$O$_7$ also shows a very important difference from the bilayer Cu oxides: the bond Ni-O-Ni appears equally robust both in-plane and in-between planes~\cite{Zhang:nc24}. Therefore, studying a single Ni–oxide plane is {\it insufficient} to capture the full physics of this compound, because the two planes in each bilayer are strongly interlocked. This idea was proposed in early theoretical works, leading to the important concept of ``dimers'' in bilayers~\cite{Zhang:prb23,Zhang:nc24}. These dimers arise from the abstraction based on focusing on just two Ni atoms, one above the other, one in each plane. Even without any Hubbard or Hund interaction, because of the two-site nature of the problem, the $d_{3z^2-r^2}$ orbitals form bonding and antibonding states due to the strong interlayer hopping, as illustrated in Fig.~\ref{fig2}(d).

Various theoretical studies have found that the electronic hopping between layers, as in the tight binding sense, is robust, larger by about $30 \%$ than the electronic hopping in plane~\cite{Luo:prl23,Zhang:prb23,Zhang:nc24,Liao:arxiv,Cao:arxiv,Lechermann:arxiv,Geisler:npjqm-24}. This observation further supports the notion that dimers could be a foundational building block to start the understanding of these compounds. Notably, early theoretical work in the context of $t-J$~\cite{Dagotto:prb92} or, in more detail, Hubbard~\cite{Maier:prb11,Kuroki:prb17} models using one orbital already hinted at the notion that, as the interlayer hopping increases and the system is doped with holes, superconductivity could emerge from the theoretical perspective. Revisiting these foundational studies may provide valuable guidance for theoretical investigations of the current two-orbital Ni oxides, such as La$_3$Ni$_2$O$_7$.

As a partial summary, two factors are particularly important: (1) the theoretical description of La$_3$Ni$_2$O$_7$ involves two Ni orbitals, and (2) requires two coupled Ni oxide planes, due to its robust bilayer nature. Colloquially, these two factors do not merely multiply by 4 the effort in theoretical approaches, but sometimes, as in computational work, the factor 4 may ``go to exponents''. Studying La$_3$Ni$_2$O$_7$ is truly a grand challenge for theorists.

\subsection{Strange-metal behavior in bilayer La$_3$Ni$_2$O$_7$}
\label{subsec_3-2}
\vspace{4pt}

Another important issue is whether La$_3$Ni$_2$O$_7$ is, or is not, strongly electronically correlated, like the cuprates are widely believed to be. Recent optical spectroscopy measurements indicate that La$_3$Ni$_2$O$_7$ features strong electronic correlations, suggesting this system maybe in the proximity of a Mott phase, as discussed in Refs.~\cite{Liu:prb2025corr,Liu:nc24,Xu:prb2025}. The presence of correlation manifests in cuprates via a variety of properties~\cite{Dagotto:rmp94,Scalapino:rp95}, but two prominent issues are (i) the existence of robust antiferromagnetism very close to the superconducting phase and (ii) the “strange metal” behavior of the resistivity at optimal doping, with a linear in temperature behavior. These two properties are not present in canonical phonon-driven BCS superconducting materials. We will later address the matter of magnetism, but now let us focus on ``strange metallicity'' in nickelates.

\begin{figure}
\centering
\includegraphics[width=0.48\textwidth]{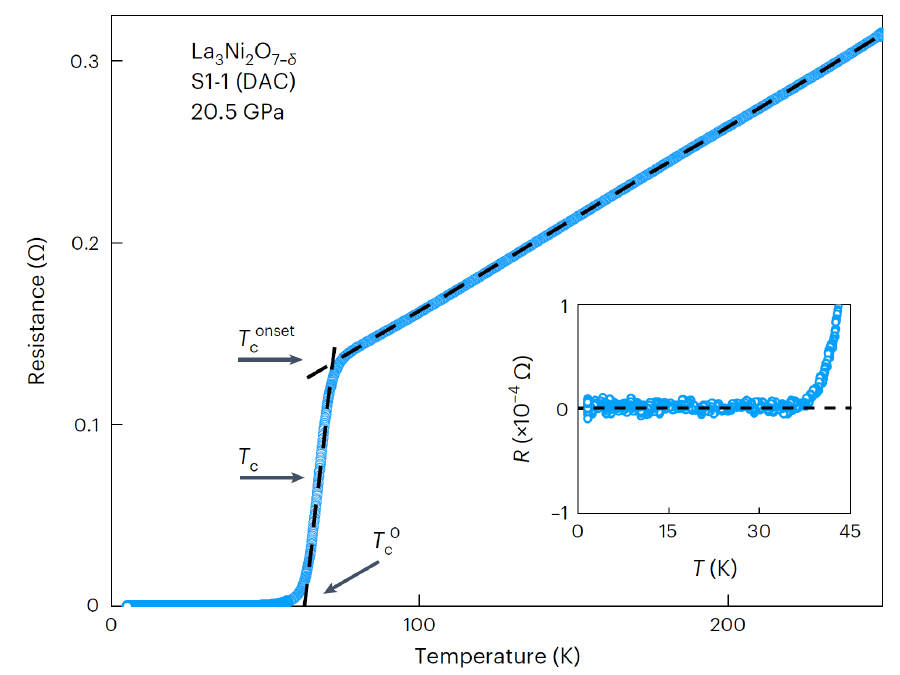}
\caption{Resistance vs. temperature for a sample of La$_3$Ni$_2$O$_7$. Reprinted with permission from Springer Nature Customer Service Centre GmbH: Springer Nature, Nature Physics, ~\cite{Zhang:arxiv-exp}, Copyright (2024).}
\label{fig3}
\end{figure}

Figure~\ref{fig3} contains the resistivity of La$_3$Ni$_2$O$_7$ at its optimal pressure~\cite{Zhang:arxiv-exp}. The onset of superconducting transition is at $T_c \sim 80$ K, while the true drop to zero of the resistance occurs at about 40 K. It is widely believed that as samples continue to improve, the onset temperature will eventually become the true $T_c$. At a lower temperature, a state of zero resistance is reached in this sample, as shown in the inset of Fig.~\ref{fig3}. Returning to our discussion of strong correlations, note the nearly perfect linear resistivity behavior in $\rho$ vs $T$ in Fig.~\ref{fig3}, which is reminiscent of the cuprates at optimal hole doping. Because this behavior also appears in the cuprates, which are widely believed to be strongly correlated, then from this result, one may conclude that the bilayer nickelates are strongly correlated as well. Note, as a word of caution, that the reason for the linear resistivity in the cuprates remains controversial. The second aspect related to strong correlations, i.e., magnetism, is still under discussion in the nickelates, and it is likely that a substantial improvement in sample quality will be needed to close this discussion. More details on this point will be given in Sec.~\ref {subsec_3-4}. Furthermore, Ref.~\cite{Zhang:arxiv-exp} is also the first observation of zero resistance in high-pressure La$_3$Ni$_2$O$_7$, followed by the confirmed Meissner effect~\cite{Li:NSR25}, establishing the true superconducting state in bilayer La$_3$Ni$_2$O$_7$.

\subsection{Electronic structure of bulk La$_3$Ni$_2$O$_7$}
\label{subsec_3-3}
\vspace{4pt}

\begin{figure*}
\centering
\includegraphics[width=0.88\textwidth]{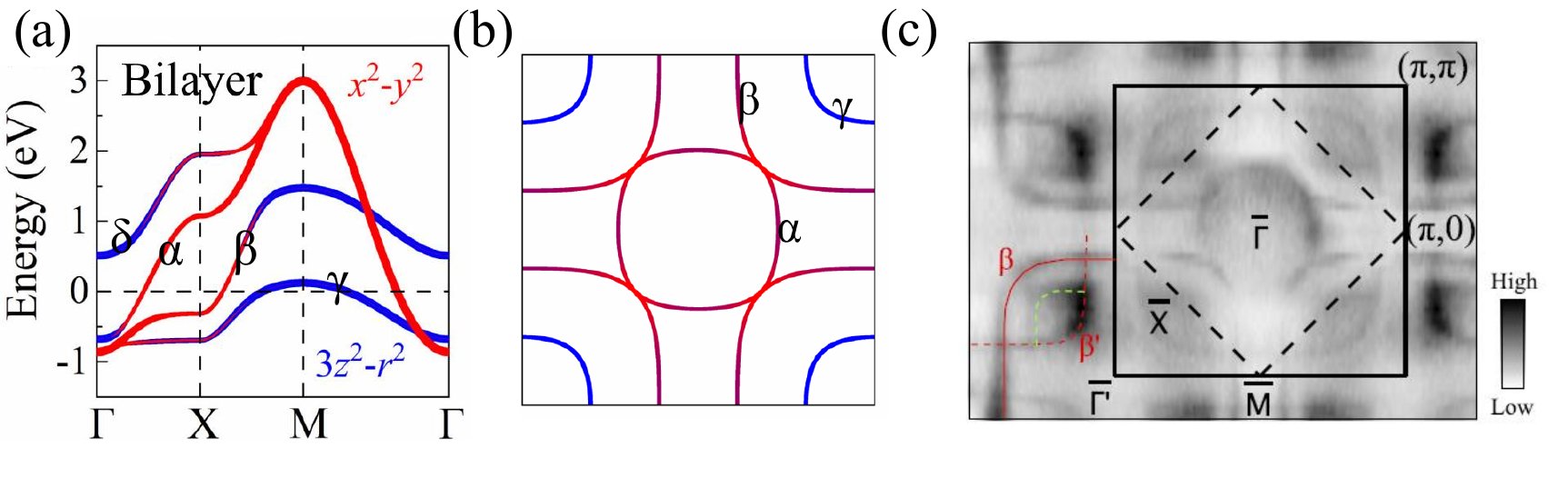}
\caption{(a) Tight-binding band structure and (b) Fermi surface of
bilayer La$_3$Ni$_2$O$_7$ under high pressure. Reprinted with permission from~\cite{Zhang:prl24}, Copyright (2025) by the American Physical Society. (c) ARPES at ambient pressure showing the presence of the predicted $\alpha$ and $\beta$ bands for La$_3$Ni$_2$O$_7$. Reprinted with permission from Springer Nature Customer Service Centre GmbH: Springer Nature, Nature Communications, ~\cite{Chen:arxiv2024}, Copyright (2025).}
\label{fig4}
\end{figure*}

Using {\it ab initio} density functional theory (DFT) methods, theoretical works by several groups provided results for
the band structure of La$_3$Ni$_2$O$_7$  in the early stages of research
on this system~\cite{Luo:prl23,Zhang:prb23,Zhang:nc24,Liao:arxiv,Cao:arxiv,Lechermann:arxiv,Geisler:npjqm-24}.
Based on the obtained hopping parameters from DFT bands under high pressure, the tight-binding band of
the two $e_g$ orbitals is displayed in Figs.~\ref{fig4}(a, b), along with the corresponding Fermi
surface~\cite{Zhang:prl24}. This result is based on the simplifying assumption of a unit cell
containing only two Ni atoms; given that each Ni contributes two relevant orbitals, a total of four bands is obtained.

Note the presence of a nearly fully occupied $\gamma$ band in Fig.~\ref{fig4}(a). The corresponding hole $\gamma$ pocket at M is an important focus of attention in the theoretical studies of bilayer nickelates. Early calculations showed that, within the random phase approximation (RPA), its presence was correlated with pairing tendencies in the $s^{\pm}$ channel, the same channel as in iron-based high-$T_c$ superconductors~\cite{Dai:np12} (more details of $s^{\pm}$ pairing will be discussed in Sec.~\ref {subsec_3-5}). Conversely, the absence of the $\gamma$ pocket was correlated with the suppression of pairing tendencies, as discussed in Ref.~\cite{Jiang:arxiv12}. In particular, previous DFT calculations also found that the $\gamma$ pockets are absent in the $Amam$ phase at ambient conditions because the $\gamma$ band is entirely below the Fermi energy~\cite{Sun:Nature23,Zhang:prb23}. Consistently, reduced RP bilayer La$_3$Ni$_2$O$_6$, which does not exhibit superconductivity at both ambient and high pressures~\cite{Liu:scpma}, also does not have the $\gamma$ pockets~\cite{Zhang:prb24}.

The absence of the $\gamma$ pocket was also confirmed by angle-resolved photoemission experiments (ARPES) at ambient pressure~\cite{Yang:arxiv09}, as shown in Fig.~\ref{fig4}(c).  While the $\alpha$ and $\beta$ portions of the Fermi surface are clearly observed, the $\gamma$ pocket is absent, consistent with theoretical predictions for the ambient pressure $Amam$ phase.

\subsection{Magnetism of bulk La$_3$Ni$_2$O$_7$}
\label{subsec_3-4}
\vspace{4pt}

\begin{figure*}
\centering
\includegraphics[width=0.88\textwidth]{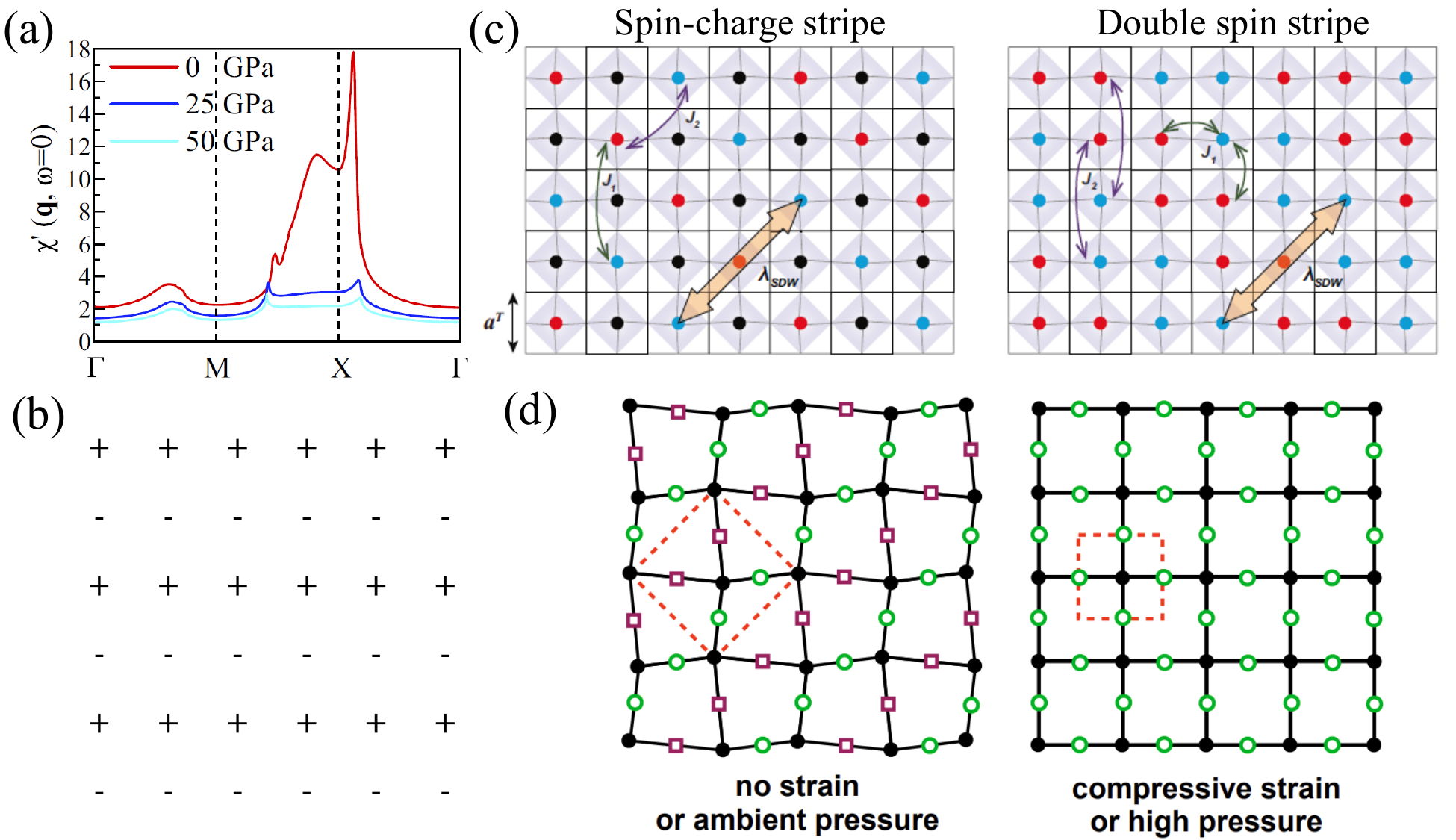}
\caption{(a) Magnetic susceptibility for the two-orbital Hubbard-Hund model, using the RPA technique. Reprinted from~\cite{Zhang:nc24}, Nature Communications, Creative Commons CC BY license. (b) Sketch of the spin patterns in real space for ${\bf q}$ = $(0, \pi)$ (or $(\pi, 0)$ when rotated by $\pi$/2) using a $6\times6$ cluster. (c) Interpretation of RIXS results for La$_3$Ni$_2$O$_7$ in the $Amam$ phase (ambient pressure) showing two arrangements compatible with the data: a spin-charge stripe order with $J_1S = 3.55 \pm 0.04$ meV, $J_2S = 2.01 \pm 0.01$ meV, and $J_zS = 67.18 \pm 0.89$ meV, and a double spin stripe with $J_1S = 0.00 \pm 0.01$ meV, $J_2S = 4.13 \pm 0.04$ meV, and $J_zS = 73.89 \pm 0.71$ meV. In the figure, blue, red, and black circles represent spin up, spin down, and spinless sites, respectively. Reprinted by permission from Springer Nature Customer Service Centre GmbH: Springer Nature, Nature Communications, ~\cite{Chen:nc2024}, Creative Commons CC-BY-NC-ND. (d) In-plane lattice distortions in the $Amam$ (left panel) and $Fmmm$ (right panel) phases. The black, red, and black represent Ni, O below the Ni-plane, and O above the Ni-plane, respectively. Reproduced from~\cite{Wang:arxiv2025-09}, CC BY 4.0.}
\label{fig5}
\end{figure*}

Within the RPA framework, magnetic tendencies in the high-pressure phase of La$_3$Ni$_2$O$_7$ can be inferred by analyzing the magnetic susceptibility and the locations of its dominant peaks. Figure~\ref{fig5}(a) shows results displaying the magnetic susceptibility from Ref.~\cite{Zhang:nc24} for various pressures. Here, the calculations at 0, 25, and 50 GPa use lattice parameters for the $Fmmm$ phase using fully relaxed lattice spacings and atomic positions. Notably, at ambient pressure for the high-pressure phase, a dominant sharp peak appears near X = $(\pi,0)$ and its symmetry-related counterpart $(0,\pi)$, similar to another theoretical work~\cite{Liu:prb25}.

For simplicity, if we assume that the peak occurs exactly at X, the corresponding real-space pattern can be readily determined, as illustrated in Fig.~\ref{fig5}(b). It contains lines of spin aligned ferromagnetically along one direction and staggered antiferromagnetically along the other direction, similar to those in Fe-based superconductors~\cite{Fernandes:prb10,Liang:prl13}. In iron-based superconductors, however, this magnetic state arises naturally due to the position of Se, P, or Te at the center of the Fe-based plaquette. This allows the electronic hopping mediated by Se, for example, to transit through the center of the plaquette with Se acting as a bridge, making nearest-neighbor (NN) and next-nearest-neighbor (NNN, along the plaquette diagonal) hoppings of comparable strength. Eventually, in the strong coupling limit, this competition naturally leads to the $(\pi,0)$ order since the corresponding Heisenberg model features NN and NNN couplings of similar value. In nickelates, however, the situation is more subtle, and the origin of the ``magnetic stripe'' dominance from a Heisenberg perspective remains under debate. A recent mechanism proposed for understanding magnetic insulating ferromagnets may play a key role in stabilizing the magnetic stripe state in these Ni oxides~\cite{Lin:prl21,Lin:prb22,Lin:cp}.

At ambient pressure, in the distorted $Amam$ phase, resonant inelastic X-ray scattering (RIXS) experiments were performed and interpreted in terms of ``stripe states'', this time oriented diagonally rather than vertically or horizontally. RIXS is well-suited for such studies because it is surface-sensitive and requires only small samples. In contrast, inelastic neutron scattering is more challenging, as it demands larger samples, and no clear results were available at the time this review was prepared. Previous elastic and inelastic neutron scattering studies on a polycrystalline sample of La$_3$Ni$_2$O$_{7-\delta}$ found that long-range magnetic order was absent down to 10 K~\cite{Xie:SB}.

Returning to RIXS, the experimental data can be interpreted as consistent with two possible arrangements, as illustrated in Fig.~\ref{fig5}(c): diagonal spin-charge stripes, which include spins up, spins down, and spinless sites (corresponding to the absence of electrons, i.e., holes), and diagonal double spin stripes, which contain only spins up and down, with no ``rivers of holes''. Both configurations are compatible with the experimental observations~\cite{Chen:nc2024}. This ($\pi/2$, $\pi/2$) possible spin pattern was also observed at $T \sim 160$ K in the La$_3$Ni$_2$O$_{7-\delta}$ film samples~\cite{Ren:cp-erxs} at the Ni L absorption edge using resonant X-ray scattering measurements, which is complementary to the RIXS study revealing the exchange energy. The fitted coupling values for both states indicate that the interlayer magnetic interactions are much stronger than the intralayer couplings, as determined from a fitting procedure using a simple Heisenberg model~\cite{Chen:nc2024}. For the double spin stripe, they obtained the coupling values with $J_1S = 0.00 \pm 0.01$ meV, $J_2S = 4.13 \pm 0.04$ meV, and $J_zS = 73.89 \pm 0.71$ meV. Very recently, another RIXS experiment also arrived to the double-stripe Heisenberg description, yielding representative exchange parameters
with $J_1S = \sim -8.0 $ meV, $J_2S = \sim -3.7$ meV, and $J_zS = 44.4$ meV for Sr-doped bilayer thin films~\cite{Zhong:RXIS}.

The double-stripe structure can also be interpreted as an E-phase according to Goodenough's notation for antiferromagnetic patterns. This E-phase has been discussed previously in another family of oxides, the manganites, which exhibit colossal magnetoresistance and are widely believed to host active Jahn-Teller modes~\cite{Dagotto:rp}. Essentially, the E-phase framework replaces the notion of ``double stripes'' with a zigzag spin pattern. We favor the E-phase interpretation in nickelates because the NN electron hopping is much stronger than hopping along the plaquette diagonals, which would be the electronic path in truly diagonal stripes. Notably, in manganites, these zigzag chains have all spins aligned in the same direction, a feature that also appears in the RIXS analysis of nickelates. An interesting hypothesis is proposed in Ref.~\cite{Wang:arxiv2025-09} for this stripe. A top-down view of the Ni oxide layers in the bilayer $Amam$ phase is shown in the left panel of Fig.~\ref{fig5}(d). Focusing on the green bonds, one can see that they form a zigzag pattern reminiscent of the E-phase in manganites. This suggests that the spins may be adapting to a preexisting lattice distortion, resulting in zigzag chains consistent with the RIXS results. This scenario is also supported by theoretical calculations based on the distorted $Amam$ phase at ambient pressure~\cite{Zhang:prb2025,LaBollita:prm,Ni:qm25}. Spefically,  they find an interesting E-phase pattern with alternating Ni$^{2+}$ and Ni$^{3+}$, which is intermediate between the single charge-spin stripe and the double-spin-stripe proposed by RIXS. Moreover, the interwined spin and charge instability was also discussed in theoretical works~\cite{Jiang:scpma25,Foyevtsova:prb25,Leonov:prb25,Chen:prb25-stripe,Oh:prb25,Oh:prb26,Wu:scpmma2024,Steffen:PRB2024}, where the origin of the small spin moments on Ni sites is believed to related to the electron delocalization over molecular orbitals involving multiple Ni and O sites~\cite{Jiang:scpma25,Foyevtsova:prb25}.

Although many other experiments using various techniques have investigated the magnetic properties of the $Amam$ phase of La$_3$Ni$_2$O$_7$, its exact magnetic structure has not yet been fully established and remains under debate. Muon spin relaxation (${\mu}$SR) studies of polycrystalline La$_3$Ni$_2$O$_{6.92}$ suggests a spin-charge stripe state at ambient conditions~\cite{Chen:prl24}. This state is also supported by the nuclear quadrupole resonance (NQR) data on polycrystalline samples~\cite{Yashima:jsjp}. However, the anisotropic splitting in the nuclear magnetic resonance (NMR) spectroscopy of $^{139}$La suggests the formation of a possible double spin stripe as well, with magnetic moments aligned along the $c$-axis~\cite{Dan:arxiv2024}. Recently, neutron scattering experiments on polycrystalline samples have proposed spin-stripe structures characterized by two wave vectors, ${\bf q_1} = (\pi/2, \pi/2, 0)$ and ${\bf q_2} = (\pi/2, \pi/2, \pi)$, by using symmetry-allowed models of spin structures testing against the neutron data~\cite{Plokhikh:arxiv2025}. Thus, further inelastic neutron scattering studies on more homogeneous single crystals are required to determine the intrinsic magnetic structure of the $Amam$ phase. Moreover, experimental measurements on La$_3$Ni$_2$O$_7$ under pressure are also needed to clarify the relationship between the magnetic state and the superconducting phase, although such experiments remain challenging.

\subsection{Pairing symmetry of bulk La$_3$Ni$_2$O$_7$}
\label{subsec_3-5}
\vspace{4pt}

DFT calculations found that electron–phonon coupling alone is not sufficient to induce superconductivity in La$_3$Ni$_2$O$_7$ under pressure~\cite{Yi:prb2024,Ouyang:qm,Zhan:prl25}. This suggests that the spin fluctuations play a key role for the high $T_c$. However, superconductivity may be enhanced through the interplay between electron–phonon coupling and strong electronic correlations in a positive loop~\cite{Zhan:prl25}.

\begin{figure*}
\centering
\includegraphics[width=0.88\textwidth]{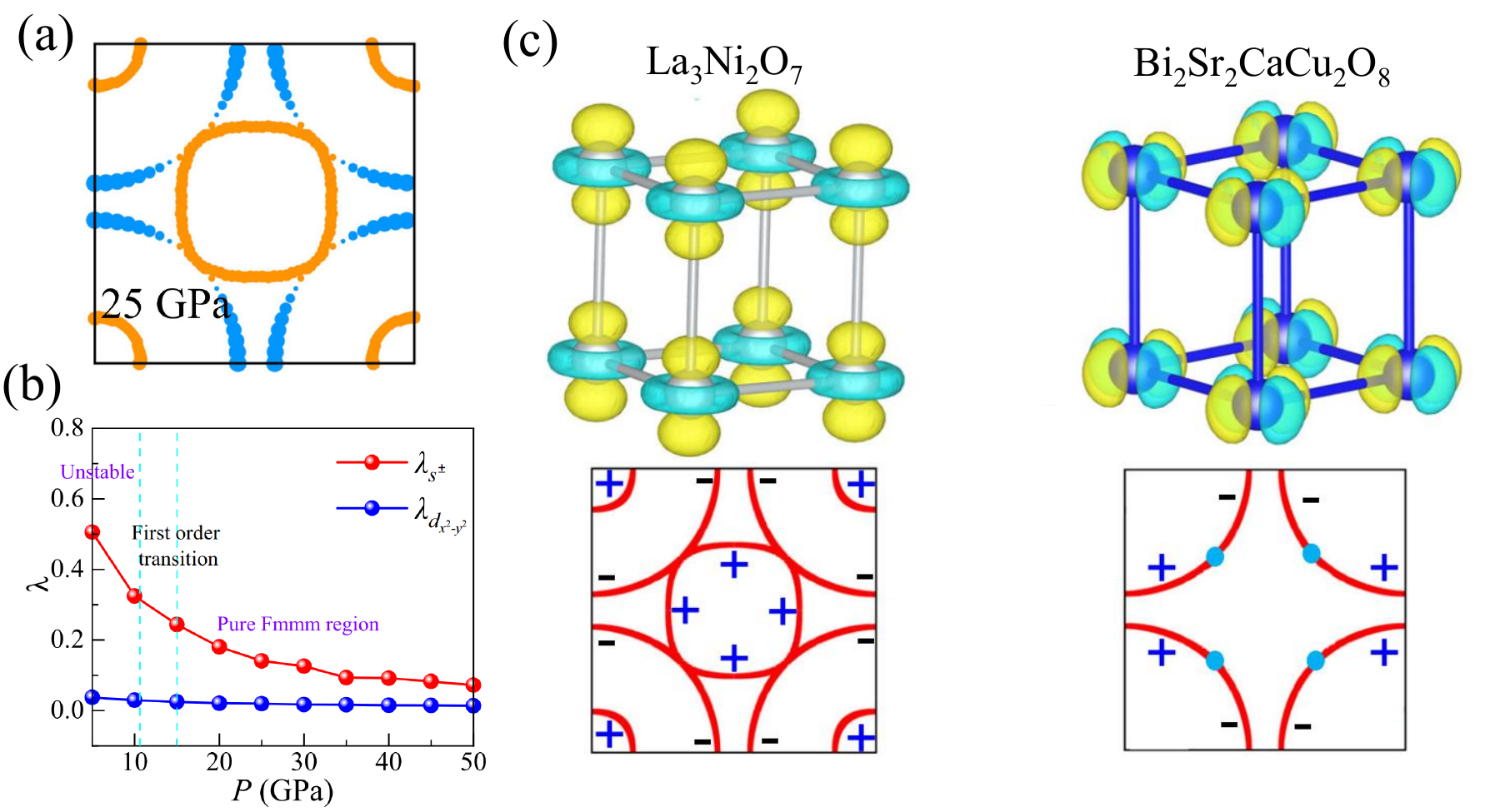}
\caption{(a) Illustration of the $s^{\pm}$ pairing state found using the RPA technique based on the standard Fermi surface of La$_3$Ni$_2$O$_7$ at high pressure by using two colors to denote the + and – signs of the superconducting order parameter. (b) The pairing strength $\lambda$ from RPA is displayed for different pressures. All results are based on the $Fmmm$ phase. (c)  Contrast between bilayer La$_3$Ni$_2$O$_7$ and Bi$_2$Sr$_2$CaCu$_2$O$_8$. Blue and yellow tones are the two signs of the orbital wave function. Reprinted from~\cite{Zhang:nc24}, Nature Communications, Creative Commons CC BY license.}
\label{fig6}
\end{figure*}

Within RPA, the magnetic susceptibility can be used as input to formulate predictions for the dominant superconducting pairing channels, under the assumption that magnetism drives superconductivity. Figure~\ref{fig6}(a) shows the Fermi surface, with two colors indicating the positive and negative signs of the RPA-derived superconducting order parameter on the different Fermi sheets and pockets (e.g., orange for + and blue for –, or vice versa)~\cite{Zhang:nc24}. Although the state is invariant under a $90^\circ$ rotation and is therefore classified as an $s$-wave, the presence of sign changes in the order parameter means that the simple ``s-wave'' terminology, commonly used in phonon-mediated BCS superconductors, is not sufficient. Additional information is required to capture this structure. Consequently, the state illustrated in Fig.~\ref{fig6}(a) is denoted as $s^{\pm}$. Notably, several independent studies based on DFT combined with RPA or other techniques have reached the same conclusion~\cite{Yang:arxiv,Liu:arxiv,Qu:prl,Lu:prl,Tian:prb24,Maier:arxiv25,Zhang:prb23-rare,Chen:scpma-26,Huang:arxiv,Luo:npjqm24}, namely that the $s^{\pm}$ channel mainly originated from interlayer coupling is dominant in this system. However, some theoretical studies have also suggested that intralayer interactions drive $d$-wave pairing in high-pressure bulk La$_3$Ni$_2$O$_7$~\cite{Lechermann:arxiv,Fan:arxiv23,Heier:prb,Braz:arxiv25,Jiang:PRL2024,Liu:arxiv2023}.

Moreover, some works~\cite{Qu:prl,Oh:PRB2023} have also argued that the interlayer $d_{3z^2-r^2}$ superexchange coupling can be effectively transferred to the $d_{x^2-y^2}$ orbitals via Hund’s coupling, with electrons in the $d_{x^2-y^2}$ orbitals playing the dominant role in superconductivity. It is important to emphasize that RPA is intrinsically a weak-coupling approach, since the Hubbard interaction $U$ is restricted to small values compared to the bandwidth. Increasing $U$ beyond a critical value drives the system toward magnetic instabilities within the RPA framework.

In Ref.~\cite{Zhang:nc24}, an interesting prediction was made. Within the $Fmmm$ phase of La$_3$Ni$_2$O$_7$ and using a DFT plus RPA approach,  the value of the pairing strength $\lambda$ was computed for varying pressures, as shown in Fig.~\ref{fig6}(b). While the subdominant $d_{x^2-y^2}$ channel barely changes with pressure, the dominant $s^{\pm}$ channel increases substantially as the pressure is reduced toward ambient conditions. Within the BCS approach to superconductivity -- even when using magnetism as the cause of the glue in Cooper pairs -- the critical temperature scales as an essential singularity exp$^{-1/\lambda}$. Therefore, the enhancement of the pairing strength shown in Fig.~\ref{fig6}(b) upon approaching ambient pressure suggests the possibility of a significantly higher $T_c$ than currently achieved under high pressure, provided that the $Fmmm$ phase can be stabilized at ambient conditions with the appearance of the $\gamma$ pockets.

It is also instructive to compare bilayer nickelates (at high pressure) and cuprates. Figure~\ref{fig6}(c) presents this comparison, with nickelates shown on the left and cuprates on the right. For bilayer nickelates, the cartoon illustrates the $d_{3z^2-r^2}$ orbital, which plays a central role in the dimer physics discussed earlier. The two colors indicate the opposite signs of the orbital wave function. Certainly, this cartoon should not be interpreted as implying that the $d_{x^2-y^2}$ orbital is not important; rather, it highlights that for Ni's case under high pressure, the $d_{3z^2-r^2}$ orbital appears to play the dominant role. In the lower-left panel of Fig.~\ref{fig6}(c), the previously discussed $s^{\pm}$ state is also shown, now represented using signs instead of colors. By contrast, the upper-right panel displays the well-known $d_{x^2-y^2}$ orbital of Cu, regarded as the key orbital for Cu in the $2+$ valence state, where the widely discussed Fermi surface of the copper oxides is shown in Fig.~\ref{fig6}(c). Since the superconducting order parameter changes sign under a $\pi/2$ rotation, it has $d$-wave symmetry, and the light blue dots indicate the well-known nodes observed in ARPES experiments.

In addition, another theoretical study~\cite{Qu:PRB2025} discussed the effects of Hund’s coupling and interorbital hybridization in high-pressure bilayer bulk systems, by using a tensor network approach on a two-orbital bilayer model. They found that, under pressure, the bulk system enters a Hund’s coupling–dominated superconducting regime. As the hybridization further increases with pressure, it would induce strong interorbital frustration, which would suppress superconducting correlations. Those results provide a natural explanation for the dome-like evolution (rise and subsequent suppression) of high-$T_c$ superconductivity in La$_3$Ni$_2$O$_7$ under high pressure.

In the main portion of this review, when the focus turns to the ultra-thin films of the La$_3$Ni$_2$O$_7$ compounds, we will show circumstances where the $d$-wave can be stabilized for nickel-oxide thin films as well, when compressive strain is used. This topic is under much discussion presently and is rapidly evolving. As will be discussed, it is conceivable that in Ni oxides, both channels $s$ and $d$ could manifest themselves under different setups.

To wrap up this review of the bulk bilayer nickelates, we note that relatively few studies have addressed the two-orbital Hubbard–Hund model beyond the regime of validity of DFT + RPA, and without resorting to simplified $t-J$ models. This distinction is important because, as discussed in the experimental sections of this review, increasing evidence suggests that the Ni oxides lie in an intermediate-to-strong coupling regime, with $t-J$ models strictly corresponding to the strong-coupling limit~\cite{Oh:PRB2023,Yang:PRB2024-DMRG,Tatsuya:PRB2025,Zhanga:PRL24-tj,Kakoi:PRb2024}. One example is that the $s^{\pm}$ pairing was also supported by a study using functional renormalization group (FRG) in the strong coupling limit via the multi-orbital Hubbard model~\cite{Yang:arxiv}, which is also supported by a variational Monte Carlo study~\cite{Liu:PRL2025-qmc}. Moreover, by introducing hydrostatic pressure in the itinerant picture within FRG, the same group~\cite{Jiang:PRL2025} also predicts a decreasing transition temperature versus pressure, in qualitative agreement with the experiments.

Another recent example of an unbiased approach, which in principle is valid at any Hubbard strength, is provided in Ref.~\cite{Maier:arxiv25}, where the dynamical cluster approximation (DCA) combined with a quantum Monte Carlo technique was employed. In this framework, a finite cluster of $N_c$ sites ($N_c$ = 2, 8, and 16) is self-consistently embedded in a dynamical mean field representing the rest of the lattice, mimicking the bulk crystal. A leading $s^{\pm}$ superconducting instability was found with a critical $T_c \sim 100$ K, comparable to experimental values. Because these results are nonperturbative, they provide a more robust foundation for the many studies in bulk systems based on more approximate methods, such as DFT + RPA, which also tend to find $s^{\pm}$ pairing. Although this DCA work focuses on bulk systems rather than thin films, it can also be extended to thin-film geometries in future investigations.

\subsection{Additional remarks for RP nickelates}
\label{subsec_3-6}
\vspace{4pt}

Before addressing the thin films of nickelates, the primary focus of this review, three additional aspects are worth remarking to the readers.

\begin{itemize}

\begin{figure*}
\centering
\includegraphics[width=0.8\textwidth]{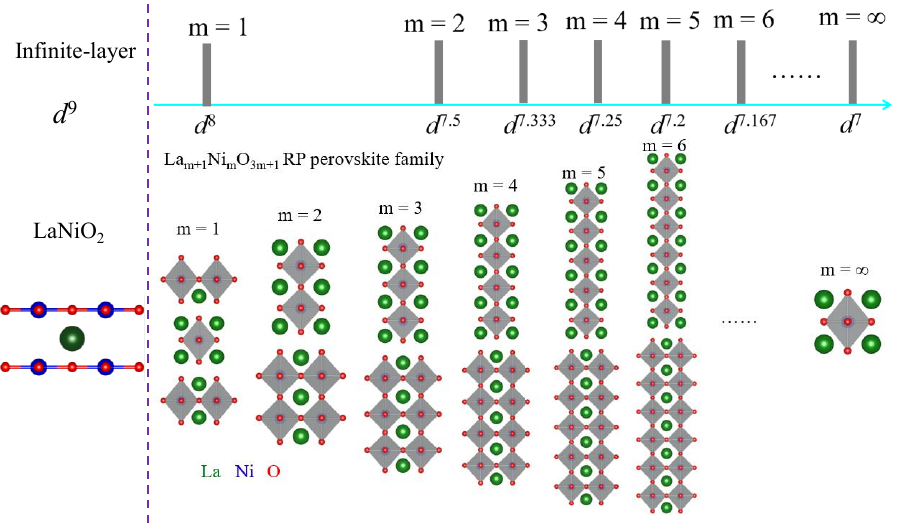}
\caption{Schematic crystal structures and electronic densities of $3d$ electrons per Ni of the RP perovskite family La$_{m+1}$Ni$_m$O$_{3m+1}$ (green = La; blue = Ni; red = O) and infinite-layer LaNiO$_2$. Reprinted with permission from~\cite{Zhang:prb25-highorder}, Copyright (2025) by the American Physical Society.}
\label{fig7}
\end{figure*}

\item  In addition to nickel oxide bilayers, there is significant interest in trilayers and even thicker multilayer systems. Figure~\ref{fig7} provides a clear overview of the various Ni valences across the RP and infinite-layer nickelates~\cite{Zhang:prb25-highorder}. Among these nickelates, three systems exhibit superconducting phases: the bilayer La$_3$Ni$_2$O$_7$ system, which is our primary focus; the infinite-layer nickelates discussed in Sec.2; and the trilayer La$_4$Ni$_3$O$_{10}$, which shows superconductivity with $T_c \sim 20-30$ K under pressure~\cite{Sakakibara:arxiv09,Li:cpl,Zhu:arxiv11,Zhang:arxiv11,Zhang:PRX15,Zhangm:prb24,Yang:prb2024-wang,Yang:arxiv2024-tri}, as illustrated in Figure~\ref{fig7}. The $T_c$ is the highest for the case of La$_3$Ni$_2$O$_7$, likely due to the robust bonding–antibonding structure discussed earlier. Interestingly, the $\gamma$ pockets were also not observed in La$_4$Ni$_3$O$_{10}$ at ambient pressure~\cite{Li:nc17} but are predicted to appear under high pressure~\cite{Zhang:prl24}. We do not review the studies related to the La$_4$Ni$_3$O$_{10}$ system here, because our focus is the thin films. Other members of the nickelate family, such as La$_2$NiO$_4$ and LaNiO$_3$, have not been observed to exhibit superconductivity experimentally~\cite{Zhang:jmst}.

\item  Regarding the density of states and the relative positions of the Ni and O bands, Ref.~\cite{Goodge:pt2025} discussed the differences between infinite-layer nickelates, cuprates, and perovskite nickelates. For the infinite-layer nickelates, it is widely believed that they fall into the Mott–Hubbard regime, where the oxygen band lies below the lower Ni Hubbard band. In contrast, for the cuprates, the oxygen band sits between the lower and upper Hubbard bands of Cu. Namely, in the case of infinite-layer nickelates, the Hubbard interaction $U$ of the Ni $3d$ orbitals is smaller than the charge-transfer energy $\Delta$ between Ni $3d$ and O $2p$ orbitals ($U < \Delta$), whereas in the cuprates, the Hubbard interaction is much larger than the charge-transfer energy ($U > \Delta$). The RP nickelates that are the focus of this review, appear to lie in an intermediate regime, with $U \sim \Delta$. The physical consequences of this observation remain an open question and warrant further investigation.

\begin{figure*}
\centering
\includegraphics[width=0.88\textwidth]{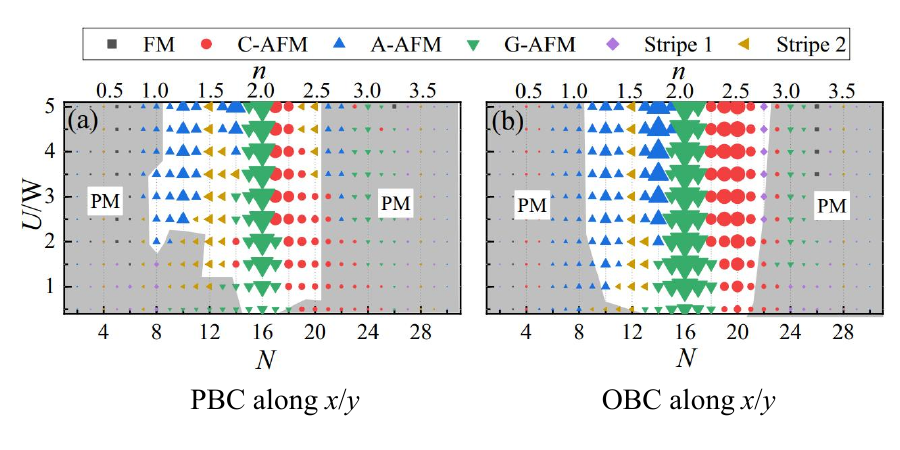}
\caption{Phase diagram of a small $2\times2\times2$ cluster solved exactly. The left panel is with periodic boundary conditions (PBCs) only along the $x$ and $y$ directions and open boundary conditions (OBCs) along the $z$ direction. The right panel is for the case where OBCs are used in the three directions.  Reprinted by permission from  Ref.~\cite{Lin:prb24}, Copyright (2024) by the American Physical Society.}
\label{fig8}
\end{figure*}

\item  Finally, it is important to note that varying the hole density provides another ``knob'' for theoretical studies of the octahedral nickelates we focus on. Because the La$_3$Ni$_2$O$_7$ system is already self-doped, adding further holes or electron doping has not been pursued extensively experimentally, contrary to what has been done for the cuprates. However, theoretical studies, such as Ref.~\cite{Lin:prb24}, have explored this systematically from the theory perspective. In that work, an electronic Hubbard-Hund model on a $2\times2\times2$ cube was solved exactly while varying the electron density, as shown in Fig.~\ref{fig8}. At $n = 1.5$, the relevant density for the La$_3$Ni$_2$O$_7$ compound, a peak in the spin structure factor $S({\bf k})$ appears at ${\bf k} = (\pi, 0)$ or $(0, \pi)$, in agreement with results from RPA calculations despite the very different nature of these approximations. Thus, the qualitative agreement at $n=1.5$ is remarkable. At $n = 2.0,$ the intuitively expected staggered antiferromagnetic order G-AFM is obtained. These findings suggest that, despite the small cluster size, exact diagonalization can yield qualitatively reliable results. The readers can easily notice the multiple competing states that are expected to be reached by further changing $n$. This route could potentially lead to interesting discoveries in the fast-developing field of Ni oxide superconductors.

\end{itemize}

\section{Superconductivity in ultra-thin-film bilayer nickel oxides}
\label{sec_4}
\vspace{6pt}

\subsection{Discovery of superconductivity in ultra-thin films of La$_3$Ni$_2$O$_7$ at ambient pressure}
\label{sec_4-1}
\vspace{4pt}

In early 2025, landmark studies reported the emergence of superconductivity at ambient pressure~\cite{Ko:nature,Zhou:nature} in thin films of the bilayer nickelate La$_3$Ni$_2$O$_7$ grown on compressively strained LaSrAlO$_4$ (LSAO) substrates using pulsed laser deposition, extending investigations that were previously limited to high-pressure bulk systems. The films are ultra-thin, typically involving 1-, 2- or 3-unit cells with each unit cell (UC) containing two Ni bilayers. This discovery has significantly advanced the field, as experiments that are not feasible under extremely high pressure can now be performed in thin films, such as ARPES.

A wide variety of efforts followed the initial discovery of superconductivity in La$_3$Ni$_2$O$_7$/LSAO thin films~\cite{Xiang:arxiv25,Ji:arxiv25-film,Lv:aps,Liu:nm25,Kumar:arxiv2603,
Osada:cp25,Bhatt:arxiv2025,Wang:arxiv2025-ele,Zhou:arxiv2025,Tarn:arxiv25,Hao:arxiv25,Li:NSR25apres,Nie:arxiv2025-apres,Shen:arxiv2025,Fan:arxiv2025,Sun:arxiv2025,Han:arxiv2026-film}. In most of these studies to date, the substrate most widely used remains LSAO, which provides in-plane compressive strain of $-2 \%$. In the last part of Sec.~\ref {sec_4-1}, we will also discuss a recent report of superconductivity in La$_3$Ni$_2$O$_7$ thin films grown on LaAlO$_3$ (LAO) without pressure, which provides less compressive strain. To date, this is the only reported case where a substrate other than LSAO successfully led to superconductivity. Consequently, the primary focus of this review will remain on the use of LSAO substrates.

\begin{figure*}
\centering
\includegraphics[width=0.88\textwidth]{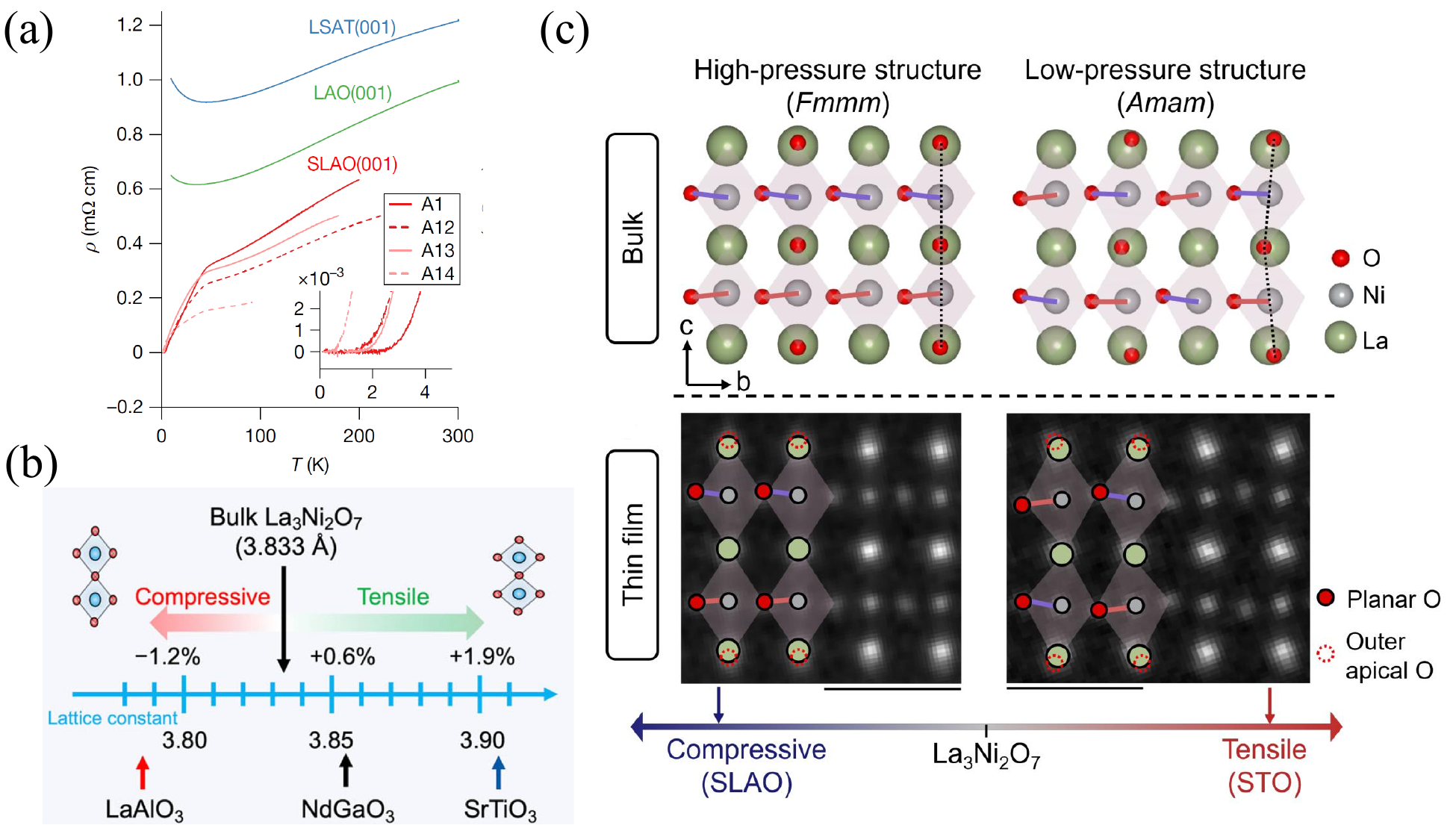}
\caption{(a) Resistivity $\rho$ vs temperature $T$ for thin films of La$_3$Ni$_2$O$_7$ grown on three different substrates. Reprinted by permission from Springer Nature Customer Service Centre GmbH: Springer Nature, Nature, ~\cite{Ko:nature}, Copyright (2025). (b) Sketch of the type of strain, compressive vs tensile, produced by the 3 substrates indicated. Reprinted by permission from Springer Nature Customer Service Centre GmbH: Springer Nature, Communications Physics, ~\cite{Osada:cp25},  Creative Commons CC-BY-NC-ND. (c) Using electron ptychography to measure the atomic-scale structural evolution of La$_3$Ni$_2$O$_7$, the atomic arrangements under strain in real space were studied. Reprinted by permission from Springer Nature Customer Service Centre GmbH: Springer Nature, Nature, ~\cite{Bhatt:arxiv2025}, Copyright (2026).}
\label{fig9}
\end{figure*}

La$_3$Ni$_2$O$_7$ films grown on the LSAO substrate present an onset of superconductivity at approximately 42 K (where ``onset'' refers to a substantial bending down of the curve) and eventually reaches the zero-resistance state at around 2 K, where a diamagnetic response is also observed, as shown in Fig.~\ref{fig9}(a). This type of behavior, namely, onset at a higher temperature followed by a drop to 0 at a much lower temperature, is typical of many superconductors in the early stages of study. It is expected that, as sample quality improves, the onset temperature will converge to the true critical temperature. In Ref.~\cite{Zhou:nature}, results similar to those in Ref.~\cite{Ko:nature} were independently reported, this time using La$_{2.85}$Pr$_{0.15}$Ni$_2$O$_7$ as partial substitution of La$^{3+}$ with Pr$^{3+}$ improves the structural purity of the samples. In Refs.~\cite{Zhou:nature,Lv:aps}, Sr diffusion from the substrate into the first unit cell of the film was observed, a topic that we will return to when discussing ARPES results.

Furthermore, superconductivity was also reported in La$_3$Ni$_2$O$_7$ films grown on STO, NdGaO$_3$ (NGO), and LAO substrates, but {\it only} when additional pressure was also applied~\cite{Osada:cp25}. Figure~\ref{fig9}(b) provides a summary of the types of strain induced by those substrates. Tensile-strain substrates that induce superconductivity are rare. To our knowledge, the only reported case is for STO as a substrate, which provides a tensile strain of $+1.9 \%$ when without additional pressure, and achieves $T_c = 10$ K only when combined with 20 GPa of applied pressure~\cite{Osada:cp25}. Applying additional pressure to the STO substrate also reduces its in-plane lattice constants.

In summary, it is far more common to find compressive strain producing a substantial resistivity reduction at an onset temperature, even without high pressure. Thus, reducing the in-plane lattice spacings of La$_3$Ni$_2$O$_7$ appears key to producing superconductivity in thin films. Note that hydrostatic high pressure reduces the lattice spacings in all three directions, whereas compressive strain achieves a similar reduction only in the two in-plane directions, namely the lattice constants $a$ and $b$. Typically, the out-of-plane lattice spacing $c$ increases when the film is grown on a compressive substrate. Thus, the superconducting states reached under high pressure or high strain are not necessarily identical, as they differ in the spacing between Ni–oxide layers, which is governed by $c$. This important distinction will be discussed in more detail from a theoretical perspective later in Sec.4.

The schematic in Fig.~\ref{fig9}(c) shows the atomic arrangements under different strains in real space using the electron ptychography. Together with in-plane compression, the strain applied to the La$_3$Ni$_2$O$_7$-material grown on LSAO transforms its $Amam$ atomic structure into the $Fmmm$ arrangement, characteristic of bulk single crystals at ambient pressure, which is the same structure obtained under high pressure in the superconducting phase. Thus, there is a clear convergence of results indicating that $Fmmm$ is indeed the fertile ground to achieve superconductivity, in both frameworks of high-pressure and high-compressive strain. Additional pressure could probably stabilize the untilted $Fmmm$ phase under STO tensile strain and induce superconductivity, albeit with a low $T_c$.

\begin{figure*}
\centering
\includegraphics[width=0.88\textwidth]{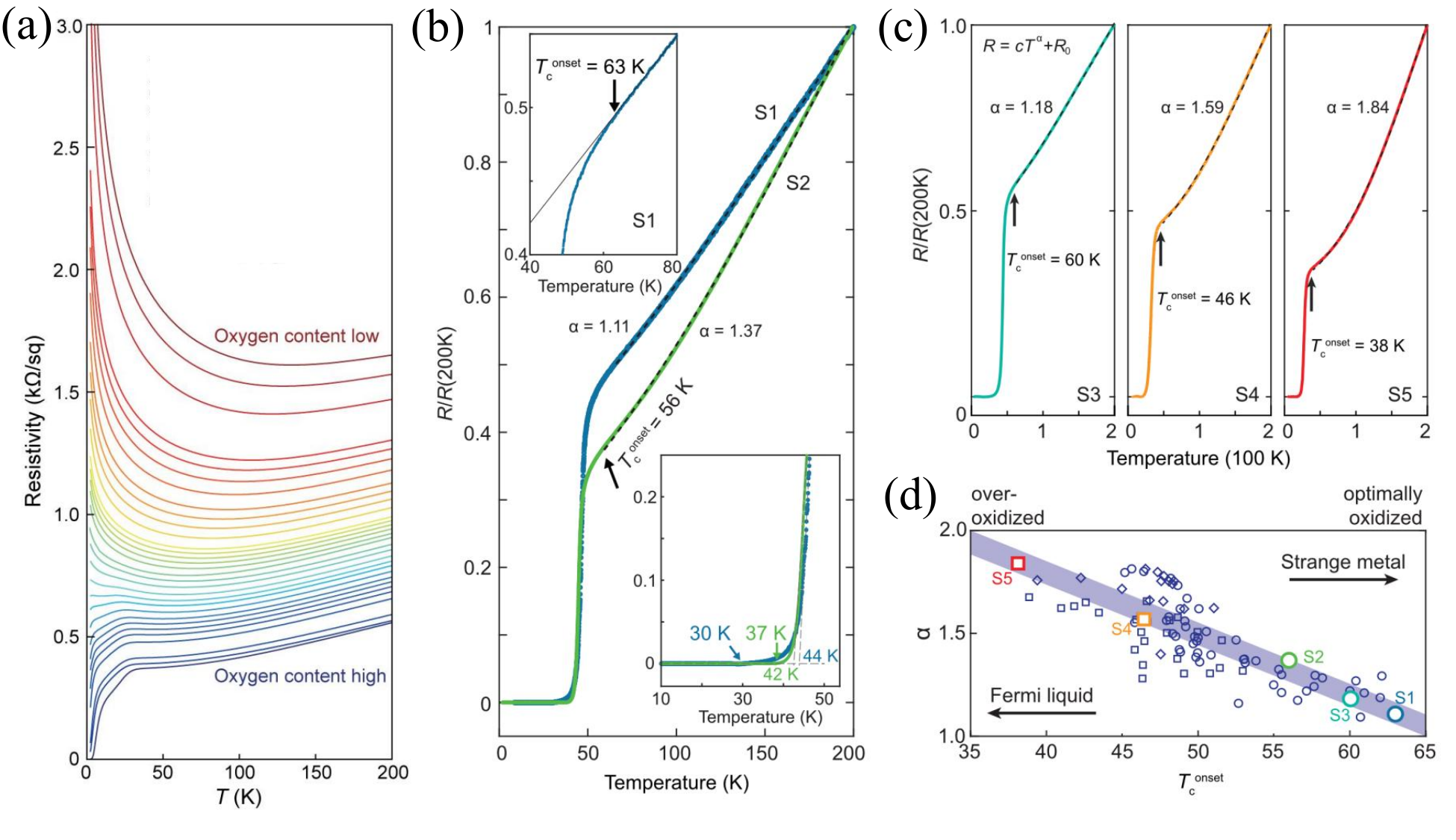}
\caption{(a) Superconductor-insulator transition in the  La$_{2.85}$Pr$_{0.15}$/LSAO interface with temperature-dependent resistivity for different oxygen contents. It is the annealing process that controls the oxygen content. Reprinted from ~\cite{Wang:arxiv2025-ele}, CC BY 4.0. (b) Recent record $T_c$ of 63 K on ultra-thin films (La, Pr)$_3$Ni$_2$O$_7$ bilayer grown on LSAO. (c) Evolution of the $\rho$ vs. $T$ curves as $T_c$ decreases. The best power fit moves from close to 1 at the highest $T_c$, to close to 2 at the lowest $T_c$. (d) Several results (90 samples, with different Pr/La ratios) from many films and oxidizations are shown for the exponent in the resistivity dependence on temperature. (b-d) are reprinted by permission from  Refs.~\cite{Zhou:arxiv2025},  National Science Review, 2026, CC BY 4.0..}
\label{fig10}
\end{figure*}

An important practical issue is the effect of hole or electron doping arising from deviations from the stoichiometric La$_3$Ni$_2$O$_7$ formula. In other words, the ``7'' in the O concentration of La$_3$Ni$_2$O$_7$ represents an approximate value effectively ``close to 7'' in practice due to the well-known difficulty of preparing samples with a precisely known oxygen content. Thus, the La$_3$Ni$_2$O$_7$ system can be labeled as ``self-doped'' from two perspectives. First, and more importantly, the number of electrons in the two active Ni orbitals is 3 per pair of Ni atoms located in different layers of the bilayer. In addition, similar to bilayer Bi$_2$Sr$_2$CaCu$_2$O$_8$, the La$_3$Ni$_2$O$_7$ system is also self-doped, owing to the fact that the oxygen content is not precisely fixed at 7. As an example, Fig.~\ref{fig10}(a) shows that in La$_{2.85}$Pr$_{0.15}$Ni$_2$O$_7$/LSAO, the resistivity decreases markedly with increasing oxygen content, evolving from a weakly insulating behavior to a metallic state, and eventually to a superconducting state~\cite{Wang:arxiv2025-ele}. This film was grown by their self-developed epitaxy method, namely, gigantic-oxidative atomic-layer-by-layer epitaxy (GAE), which is fundamentally different from PLD thermodynamically and kinetically, although both utilize pulsed lasers~\cite{Zhou:nsr-gae}. In Ref.~\cite{Lv:aps}, the optimal growth conditions were further explored, raising the onset $T_c$ to 50 K, where the film was also grown using the GAE method. This resembles the case of infinite-layer NdNiO$_2$, which becomes superconducting only after a reduction process transforms NdNiO$_3$ into NdNiO$_2$ via the removal of apical oxygens. In summary, the removal or refilling of oxygen is crucial for stabilizing the superconducting state.

For simplicity, rather than reproducing here the many subsequent record $T_c$ values achieved through improvements in thin-film growth, we instead focus on the current record holder at the time of writing [see Fig.~\ref{fig10}(b)], obtained using GAE. In Ref.~\cite{Zhou:arxiv2025}, an onset $T_c = 63$ K was recently reported for a thin film of (La, Pr)$_3$Ni$_2$O$_7$ grown on LSAO at ambient pressure. For comparison, the ambient-pressure records are 151 K for cuprates~\cite{Deng:pnas26} and 55 K for Fe-based superconductors~\cite{Ren:cpl08}. Notably, all these values, including the 63 K record for nickelate films, exceed the McMillan limit of $\sim 40$ K commonly cited for phonon-mediated superconductivity within the BCS framework.

It is important to note that, for this record-$T_c$ thin film, the slope of the $\rho$ versus $T$ curve is close to 1.
However, also note that in general, the fitting methods in different papers can be different and thus may affect the powers reported. In contrast, for the films grown under more oxidizing conditions, as $T_c$ decreases, the dominant temperature dependence of $\rho(T)$ gradually shifts toward Fermi-liquid behavior, following a $T^2$ law, as illustrated in Fig.~\ref{fig10}(c). A recent study indicates thin nickelate films could be in proximity to a strongly fluctuating ordered state, leading to non-Fermi liquid behavior under pressure~\cite{Kumar:arxiv2603}.

The Fermi-liquid behavior, $\rho \sim T^2$, has also been observed in films of La$_2$PrNi$_2$O$_7$~\cite{Liu:nm25,Hsu:nc26}.
Note that Ref.~\cite{Hsu:nc26} accesses the normal state for temperatures less than 10 K by suppressing superconductivity with a magnetic field. In Ref.~\cite{Liu:nm25} and the references cited therein, it was noted that the isovalent substitution of La by Pr suppresses competing atomic arrangements and, overall, benefits superconductivity, substantially increasing the critical current density. From the theory perspective, seemingly the ratio of La$^{3+}$ and Pr$^{3+}$ should not matter, since both have the same valence; however, in practice, partial Pr$^{3+}$ doping is important to stabilize the superconducting state. Ref.~\cite{Hsu:nc26} further pointed out that Fermi-liquid behavior in a $T_c \sim 40$ K thin film was accompanied by a large quasiparticle effective mass, $m^*/m \sim 10$, indicating a highly renormalized Fermi-liquid state. Therefore, even a quadratic temperature dependence of the resistivity does not necessarily mean a good metal. Note, however, that recent ARPES results~\cite{Sun:arxiv2025} showed the  Ni $3d_{x^2-y^2}$-derived $\alpha$ and $\beta$ bands exhibit more moderate electron correlations, characterized by a band renormalization factor of only 3-4.

In Fig.~\ref{fig10}(d), a summary of the power laws in $\rho$ versus $T^a$ for multiple samples is presented. This clearly indicates that as $T_c$ increases, the slope of the $\rho$ versus $T$ curve converges to a ``strange metal'' behavior, $\rho \sim T$, as widely observed in the cuprates at optimal doping~\cite{Dagotto:rmp94}. This suggests similarities between the superconducting state of Ni and Cu oxides regarding the normal state behavior. Importantly, this does not imply that the pairing channel must be the same, just that the normal states, which presumably contain the seed for pairing, share robust similarities.

It is also worth noting that some groups, after growing superconducting thin films, in addition subject the samples to high-pressure hydrostatic conditions (see, for example, Refs.~\cite{Osada:cp25,Li:arxiv2025-film}). The outcome is that $T_c$ of the thin films increases under pressure, in a manner similar to bulk La$_3$Ni$_2$O$_7$ samples. This provides consistency to the ensemble of experiments: it seems that high pressure, whether on bulk or thin films, always increases $T_c$.

Recent thin-film work reports the growth of relatively {\it thick} films, with thicknesses reaching up to 23.5 nm~\cite{Shi:AM26}. They observe superconductivity in a region of width 10 nm close to the interface with LSAO. Beyond this interfacial region, the atomic structure of the nickelate shifts into the trilayer form (La,Pr)$_4$Ni$_3$O$_{10}$. This observation is particularly noteworthy, as the trilayer compound is also known to exhibit superconductivity in bulk under high pressure. These results suggest that an alternative strain environment—potentially involving stronger compressive strain—may be necessary to stabilize superconductivity in trilayer thin films.

Exciting recent results report superconductivity in thin films of (La,Pr)$_3$Ni$_2$O$_7$ grown on an LAO substrate without pressure, which is the first time this happens on a substrate other than LSAO~\cite{Tarn:arxiv25}. Here, LAO provides compressive strain of $-1.2 \%$, smaller than the $-2 \%$ strain from LSAO. The films were grown using both pulsed laser deposition and molecular beam epitaxy. Notably, $T_c$ is reduced to $\sim 10$ K on LAO compared to LSAO. In both cases, the superconducting films adopt a tetragonal structure, similar to high-pressure bulk samples.

\begin{figure*}
\centering
\includegraphics[width=0.8\textwidth]{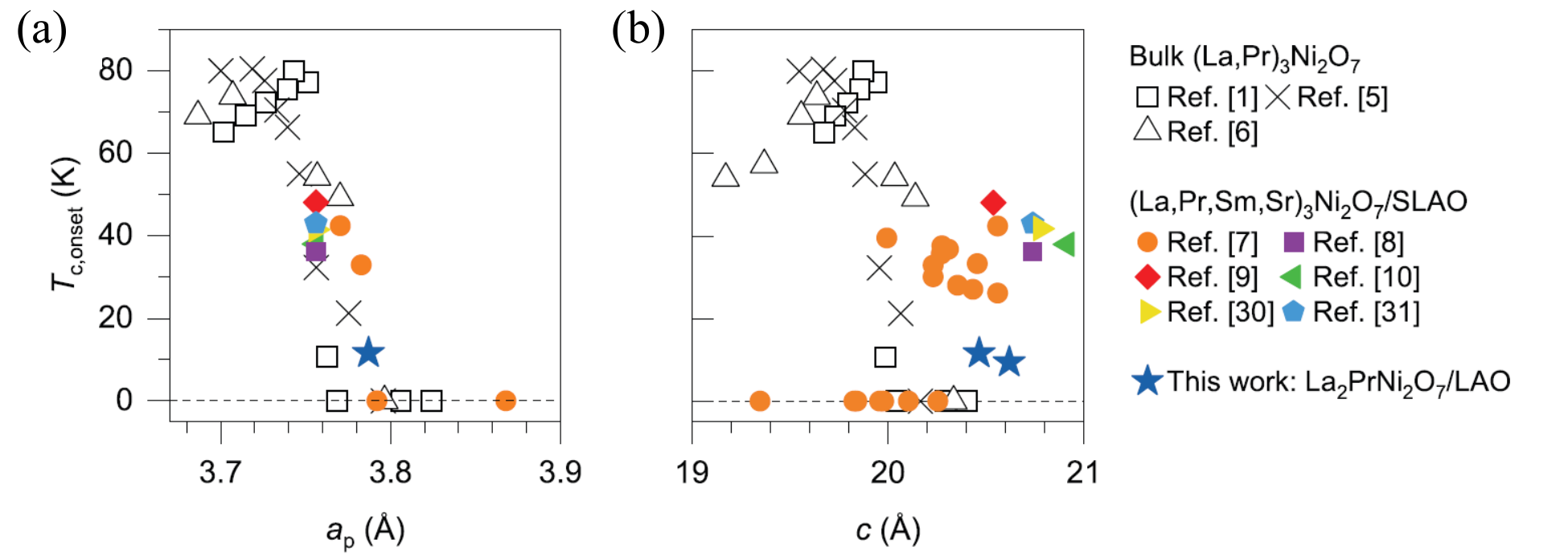}
\caption{Trend of $T_c$ vs the average in-plane lattice spacing, $a_p$, mixing results from bulk crystals at high pressure, as well as high compressive strain thin films. The common trends are remarkable, fitting into a smooth curve, unveiling a critical $a_p$ below which superconductivity emerges. (a) Trend of $T_c$ vs the lattice spacing in-plane $a_p$. (b) Trend of $T_c$ vs the lattice spacing out of plane $c$. Reprinted by permission from  John Wiley and Sons Customer Service, Advanced Materials, ~\cite{Tarn:arxiv25}, Copyright (2026).}
\label{fig11}
\end{figure*}

It is instructive to discuss similarities and differences in the trends regarding the inter- and intra-layer lattice spacings for superconductivity in nickelates. Figure~\ref{fig11}(a) shows $T_c$ as a function of the average in-plane lattice constant, $a_p$, combining data from both high-pressure bulk samples (non-colored symbols) and thin films (colored symbols). Remarkably, the resulting smooth curve demonstrates a consistent trend between high-pressure bulk and high-strain thin-film samples. There appears to be a ``critical $a_p$'' close to 3.8~\AA, below which superconductivity emerges. Notably, the smallest $a_p$, close to 3.7~\AA, has not yet been achieved in thin films, which qualitatively explains why the $T_c$ of thin films remains lower than that observed in high-pressure experiments, as shown in Fig.~\ref{fig11} (a).

Apparently, even greater compressive strain is required in thin films to raise $T_c$ to the levels achieved under high-pressure conditions. The behavior of the out-of-plane lattice spacing $c$ is different, as expected, as displayed in Fig.~\ref{fig11} (b). Under high-pressure conditions, which are essentially hydrostatic, a reduction in $a_p$ is accompanied by a simultaneous reduction in $c$. In contrast, for the thin films of interest here, Poisson-ratio effects dominate: while compressive strain reduces $a_p$, $c$ increases concurrently, at least within the range of compressive strain typically applied.

\subsection{Influence of hole doping away from stoichiometry La$_3$Ni$_2$O$_7$}
\label{sec_4-2}
\vspace{4pt}

\begin{figure}
\centering
\includegraphics[width=0.5\textwidth]{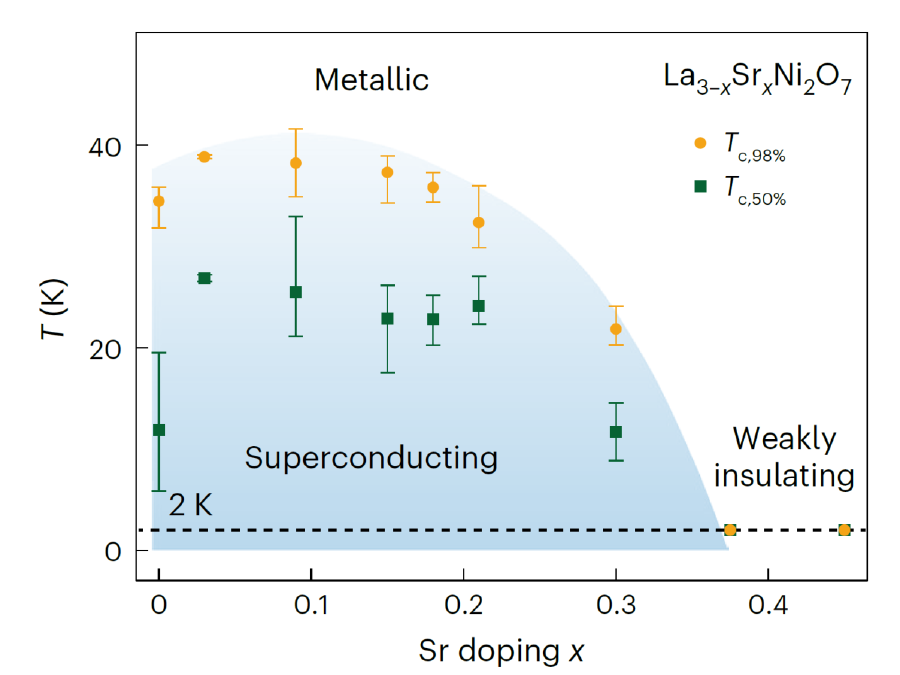}
\caption{Critical temperature $T_c$ of 3 UC thin films of La$_{3-x}$Sr$_x$Ni$_2$O$_7$ grown on LSAO. The two colors indicate different definitions of $T_c$: orange corresponds to the temperature where the resistivity reaches $98 \%$ of the extrapolated normal-state resistivity, while green corresponds to $50 \%$. Reprinted by permission from Springer Nature Customer Service Centre GmbH: Springer Nature, Nature Materials, ~\cite{Hao:arxiv25}, Copyright (2026)}
\label{fig12}
\end{figure}

Because of the self-doped nature of La$_3$Ni$_2$O$_7$, experimental studies exploring the effects of additional doping have been limited, but some results are available. For example, Ref.~\cite{Hao:arxiv25} reports the superconducting phase diagram of La$_{3-x}$Sr$_{x}$Ni$_2$O$_7$ thin films. Investigating the phase diagram under doping is particularly important in cases where, for the undoped system without pressure, the $\gamma$-pocket lies below the Fermi level and superconductivity is absent. Many theories for
La$_3$Ni$_2$O$_7$ suggests that the $\gamma$ pockets are necessary for proper Fermi-surface nesting to induce superconductivity. Thus, hole doping in this context can lift the $\gamma$-pocket above the Fermi level, potentially promoting superconductivity.

Figure~\ref{fig12} shows a superconducting phase with a peak $T_c$ at approximately $9 \%$ Sr doping for the thin-film sample~\cite {Hao:arxiv25}. Overall, critical temperatures remain fairly similar over the range $x = 0.0$ to $0.20$, beyond which $T_c$ drops rapidly. An overall conclusion of this review, at least as of its writing, is that even with the aid of Sr doping, the $T_c$ of thin films does not exceed that of high-pressure La$_3$Ni$_2$O$_7$ bulk superconductors. This limitation may be attributed to the significant elongation of the $c$-axis lattice spacing, which is inevitable in thin films. However, the same group does not observe the $\gamma$ band crossing the Fermi level in their ARPES measurements for this Sr-doped La$_3$Ni$_2$O$_7$ film sample~\cite{Sun:arxiv2025}. This puzzle of the absence of the $\gamma$ pocket still needs more work to be clarified.

Very recently, a superconducting half-dome as a function of continuously tuned oxygen stoichiometry in compressively strained bilayer nickelate thin films was also reported in Ref.~\cite{Liu:arxiv2026-dome}. It is believed that interstitial oxygen and oxygen vacancies play key roles in shaping this half-dome phase diagram. Increasing the oxygen stoichiometry gradually suppresses superconductivity from an initially optimal superconducting state, whereas decreasing the oxygen stoichiometry induces a granular superconductor-insulator transition.

\subsection{Fermi surfaces}
\label{sec_4-3}
\vspace{4pt}

The availability of superconducting thin films at ambient pressure has allowed experimentalists to carry out ARPES studies on these systems. Since ARPES probes only the sample surface, the use of thin films does not pose an obstacle. It should be emphasized that other ARPES studies of bilayer nickelate films at ambient pressure have reached different conclusions regarding the Fermi surface composition, specifically, whether the $\gamma$ pocket appears or not. This difference among different studies may also be explained by more of an issue of data interpretation rather than a real difference in experimental observation.

\begin{figure*}
\centering
\includegraphics[width=0.94\textwidth]{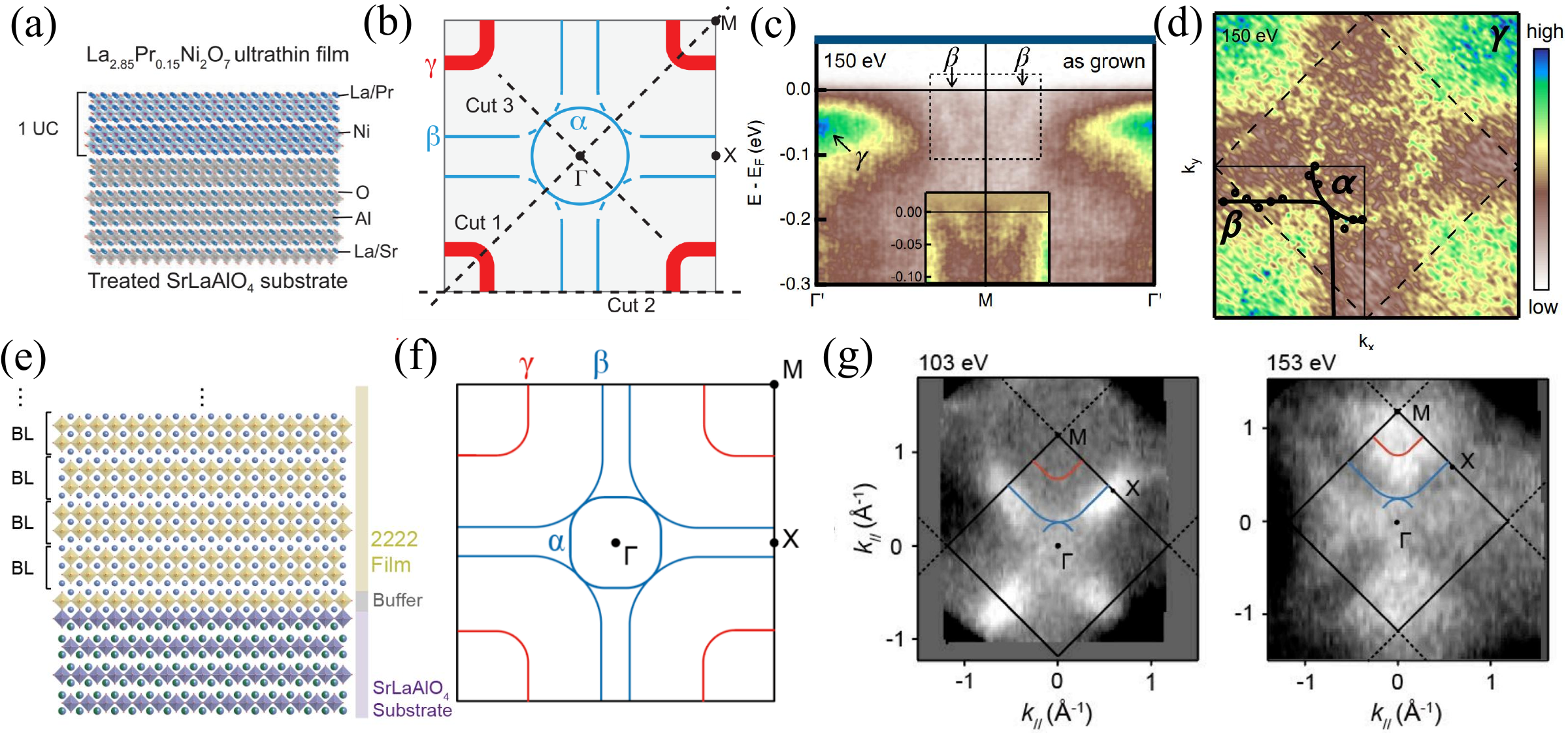}
\caption{Sketch of (a) 1 unit-cell film of (La,Pr)$_3$Ni$_2$O$_7$ on the LSAO substrate and (b) the corresponding ARPES results. The $\alpha$ and $\beta$ sheets are clearly visible, as well as the $\gamma$ pockets around the M = $(\pi, \pi)$ point. (a-b) are reprinted by permission from  Ref.~\cite{Li:NSR25apres}, National Science Review, 2025, CC BY 4.0. (c) Spectra of the as-grown MBE sample along the $\Gamma'$ to M to $\Gamma$ path. The inset shows the region near the Fermi level at the M point. (d) Fermi surface map of the same sample, using linear horizontally polarized photons with an energy of 150 eV. (c-d) are reprinted from Ref.~\cite{Wang:arxiv2025}, CC BY 4.0. Sketch of (e) a (La,Pr)$_3$Ni$_2$O$_7$ film on the LSAO substrate with buffers and (f) corresponding ARPES results. (g) Fermi surface features obtained by ARPES using different photon energies. Matrix element effects make different orbitals sensitive to different incoming photon energies; therefore, combining measurements at multiple energies is essential to capture the complete Fermi surface. (e-g) are reprinted from Ref.~\cite{Nie:arxiv2025-apres}, Copyright (2025).}
\label{fig13}
\end{figure*}

Among the earliest ARPES studies on La$_3$Ni$_2$O$_7$ thin films are those reported in Ref.~\cite{Li:NSR25apres}. Because we will focus later on the most recent ARPES results, the discussion of this early work is kept brief. Figures~\ref{fig13}(a-b) show the findings for the 1 UC thin-film sample of (La, Pr)$_3$Ni$_2$O$_7$ grown on LSAO. The Fermi surface topology contains the expected $\alpha$ and $\beta$ sheets, as well as a $\gamma$ pocket around the M point of the Brillouin zone, dominated by the Ni $d_{3z^2-r^2}$ orbitals. It was found that conduction occurs in the first unit cell near the surface. The authors of Ref.~\cite{Li:NSR25apres} suggested that Sr diffusion or oxygen loss may have induced this conduction, with overall hole doping lowering the Fermi level and rendering the $\gamma$ pocket visible. The estimated hole doping is approximately $\sim 20 \%$ per Ni relative to the non-superconducting parent compound. Notably, the features around the M point were diffuse but clearly discernible.

In Ref.~\cite{Wang:arxiv2025}, it was reported that the $\gamma$ pocket is located approximately 70 meV below the Fermi level, leading to the absence of the $\gamma$ pocket, as displayed in Figures~\ref{fig13}(c-d). A similar conclusion of the absence of $\gamma$ pocket was also made by another ARPES study~\cite{Sun:arxiv2025}. This observation has significant implications for theories that consider only the Fermi surface states for pairing, such as RPA, where the structure of the Fermi surface is central. If the $\gamma$ band does not cross the Fermi level, then this band will not contribute to pairing. However, in more sophisticated treatments that fully consider the dynamics of the pairing interaction, a band that is away from the Fermi level but within the dynamic range of the interaction will still contribute to pairing, as found in theoretical studies in the context of iron-based superconductors \cite{linscheid:prl16, mishra:scirep16, Maier:PRB19, Matsumoto:JPSJ20}. In this case, early studies of the one-orbital Hubbard bilayer~\cite{Maier:PRB19} provide a more complex perspective because it was found that whether the band is or is not at the Fermi level does affect the dynamics of the pairing interaction, particularly the frequency dependence of the gap.  This is especially relevant for an $s^{\pm}$ gap since when a Hubbard $U$ is present, this strong repulsion can only be avoided if the superconducting order parameter changes sign in frequency.

Moreover, in intermediate- and strong-coupling regimes, an energy scale of 70 meV is relatively small compared to the full bandwidth of the two relevant Ni orbitals, which is about 3-4 eV. Therefore, for theories beyond weak coupling or simple Fermi surface restricted calculations, whether the $\gamma$  band crosses the Fermi level may have little qualitative impact; what matters is that it lies close to the Fermi energy. However, it is still unclear whether the bilayer RP nickelates should be classified as weakly, intermediate, or strongly correlated, a question that will likely be addressed in future work.

Another more recent ARPES result~\cite{Nie:arxiv2025-apres} in thin films of Ni oxides grown on a substrate with buffer layers [see Fig.~\ref{fig13}(e)] showed that
the $\gamma$ band crosses the Fermi level, as shown in Fig.~\ref{fig13}(f). Meanwhile, the onset critical temperature was found to be 43-50 K. Figure~\ref{fig13} (g) shows the Fermi surface maps measured using 103 and 153 eV photons acting on a 5-nm superconducting bilayer nickelate film. This example illustrates the fact that to unveil the entire Fermi surface, results combining photons with different incoming energies are needed. In particular, the 153 eV photons are crucial for the $\gamma$-band-like features near the M point.

\subsection{Symmetry of the order parameter}
\label{sec_4-4}
\vspace{4pt}

\begin{figure*}
\centering
\includegraphics[width=0.88\textwidth]{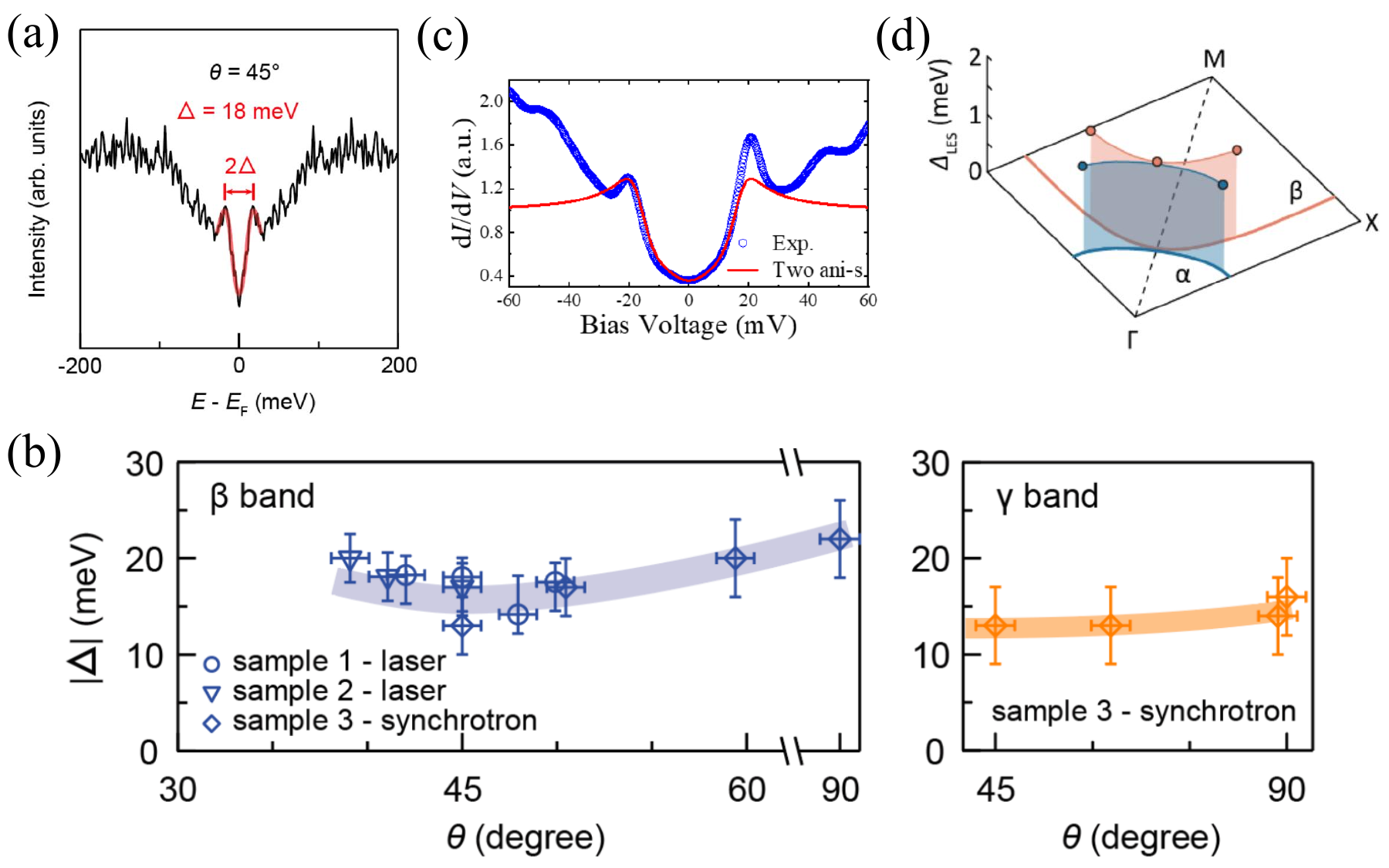}
\caption{ ARPES results for  (La,Pr,Sm)$_3$Ni$_2$O$_7$/LSAO thin films interpreted as compatible with $s$-wave pairing due to the absence of nodes.  (a) Symmetrized energy distribution curves for the Brillouin zone diagonal direction. The coherent peaks are clearly visible. Although there is a pronounced minimum at $E = E_F$, the intensity does not drop to zero. (b) Same but along the $\beta$ and $\gamma$ pockets, at an offset from the Brillouin zone diagonal. (a-b) are reprinted from  Ref.~\cite{Shen:arxiv2025}, CC BY 4.0. (c) Theoretical fitting results using the Dynes model with two gaps, and the normalized spectrum at 0.4 K. Blue points are experimental results obtained via STM. The red curve corresponds to theoretical results obtained via a phenomenological model employing a two-gap anisotropic $s$-wave order parameter.  Reprinted from Ref.~\cite{Fan:arxiv2025},  CC BY 4.0. (d) Schematic illustration of the momentum-dependent leading-edge shifts. Note the very small energy scale on the vertical axis. Reprinted from  Ref.~\cite{Sun:arxiv2025}, CC BY 4.0.}
\label{fig14}
\end{figure*}

Regarding the symmetry of the pairing order parameter, an ARPES study on superconducting bilayer (La,Pr,Sm)$_3$Ni$_2$O$_7$ thin films grown on LSAO at ambient pressure ($T_c \sim 46$ K~\cite{Shen:arxiv2025}) was interpreted as showing the absence of nodes at the Fermi surface. The LDOS exhibits a V-shaped profile (Fig.~\ref{fig14}(a)), which might suggest nodes as in a $d$-wave gap. However, with a more detailed analysis of the data, Fig.~\ref{fig14}(b) shows no nodes. Thus, it displays a nonzero gap in the studied range of the $\beta$ and $\gamma$ portions of the Fermi surface. Given the two dominant theoretical frameworks for Ni-oxide superconductors, predicting either $s^{\pm}$ or $d$-wave pairing~\cite{Fan:arxiv23,Heier:prb,Braz:arxiv25,Liu:arxiv2023}, these results are more in favor of an $s^{\pm}$ gap. Additionally, Ref.~\cite{Shen:arxiv2025} also reported that a dispersion kink indicating an electron-boson coupling was observed, which is similar to what occurs in the cuprates.

Similar conclusions regarding predominantly $s$-wave pairing were reported in Ref.~\cite{Fan:arxiv2025} using scanning tunneling microscopy (STM). The best fit to the experimental data was achieved with two anisotropic $s$-wave order parameters, as illustrated in Fig.~\ref{fig14}(c). This model involves two gap values: $\sim 19$ meV and $\sim 6–8$ meV for the thin film of 2 UC La$_2$PrNi$_2$O$_7$ with onset $T_c = 41.5$ K. Despite some sample inhomogeneity, the authors note that the highly local probing capability of STM enables the identification of regions exhibiting robust superconductivity.

However, other studies could be interpreted as compatible with nodes. In Ref.~\cite{Sun:arxiv2025}, in-situ ARPES was performed on Sr-doped superconducting thin films, specifically La$_{2.79}$Sr$_{0.21}$Ni$_2$O$_7$. The Sr doping introduces holes, and details of the doping levels are provided in Ref.~\cite{Sun:arxiv2025}. They observed a very small gap of $\sim 1–2$ meV at the Fermi momenta along the Brillouin zone diagonal, as shown in Fig.~\ref{fig14}(d). This tiny gap shift of $\sim 1-2$ meV is a leading-edge shift driven by the gap opening. Since coherent peaks were not observed, the exact superconducting gap cannot be directly extracted. Notably, the leading-edge shift is typically smaller than the true superconducting gap. Therefore, this small shift does not necessarily imply a $d$-wave symmetry. In fact, their data~\cite{Sun:arxiv2025} shows the absence of a node at the nodal point, which is consistent with other ARPES studies and favors $s^{\pm}$-pairing. Moreover, they report a $\gamma$ band 75 meV below the Fermi level, suggesting that the $d_{x^2-y^2}$ orbital is dominant. Overall, moderate band renormalizations were found. Recently, a proposal shows that Raman spectra could be a powerful symmetry-resolving probe for determining the superconducting gap in unconventional superconductors~\cite{Zhan:cpl26}.

In summary, significant work remains to be done to clarify two key questions of the superconducting Ni-oxide thin films: (i) whether the $\gamma$ pocket does or does not appear at the Fermi level, and (ii) whether the superconducting state does or does not have nodes.

\section{Theoretical studies on nickelate films}
\label{sec_5}
\vspace{6pt}

The theory of nickelate thin films is still rapidly evolving, and no consensus has been reached yet regarding the dominant pairing channel. While most studies continue to favor $s^{\pm}$ pairing, consistent with high-pressure bulk samples, others suggest a $d$-wave state.

\begin{figure}
\centering
\includegraphics[width=0.5\textwidth]{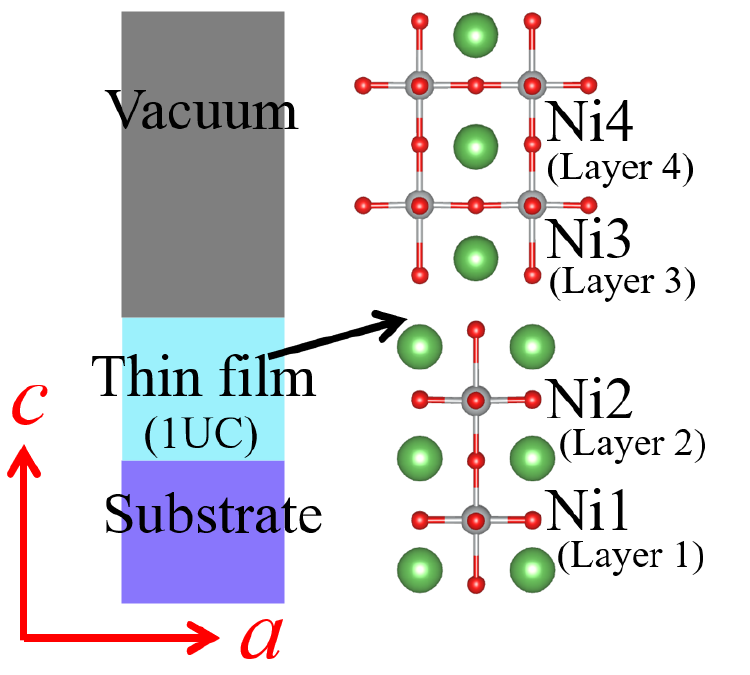}
\caption{Slab geometry widely employed for thin films in the context of {\it ab initio} calculations. The unit cell used in commonly employed codes, which assumes PBC, contains a vacuum layer of about 20~{\AA} to simulate the interface material-vacuum. The influence of the substrate is via fixing the $a$ and $b$ lattice constants, or by including a few layers of the substrate in this enlarged unit cell. For our publication Ref.~\cite{Zhang:arxiv2025-film} described below, we used 1 UC ultra-thin film. This corresponds to the 4 NiO$_2$ layers shown on the right. Reprinted by permission from  Ref.~\cite{Zhang:arxiv2025-film}, Copyright (2025).}
\label{fig15}
\end{figure}

Thin films are intrinsically more complex to study than bulk materials because they are sandwiched between a substrate and vacuum, breaking inversion symmetry along the axis perpendicular to the film~\cite{Zhang:arxiv2025-film}. In bulk calculations, periodic boundary conditions can be applied straightforwardly, but for thin films, special approaches are needed. A common solution is to enlarge the unit cell to include a vacuum layer of several angstroms (typically 20 \AA). This enlarged cell is then periodically repeated in what is known as a slab geometry, as illustrated in Fig.~\ref{fig15}.

\begin{figure}
\centering
\includegraphics[width=0.5\textwidth]{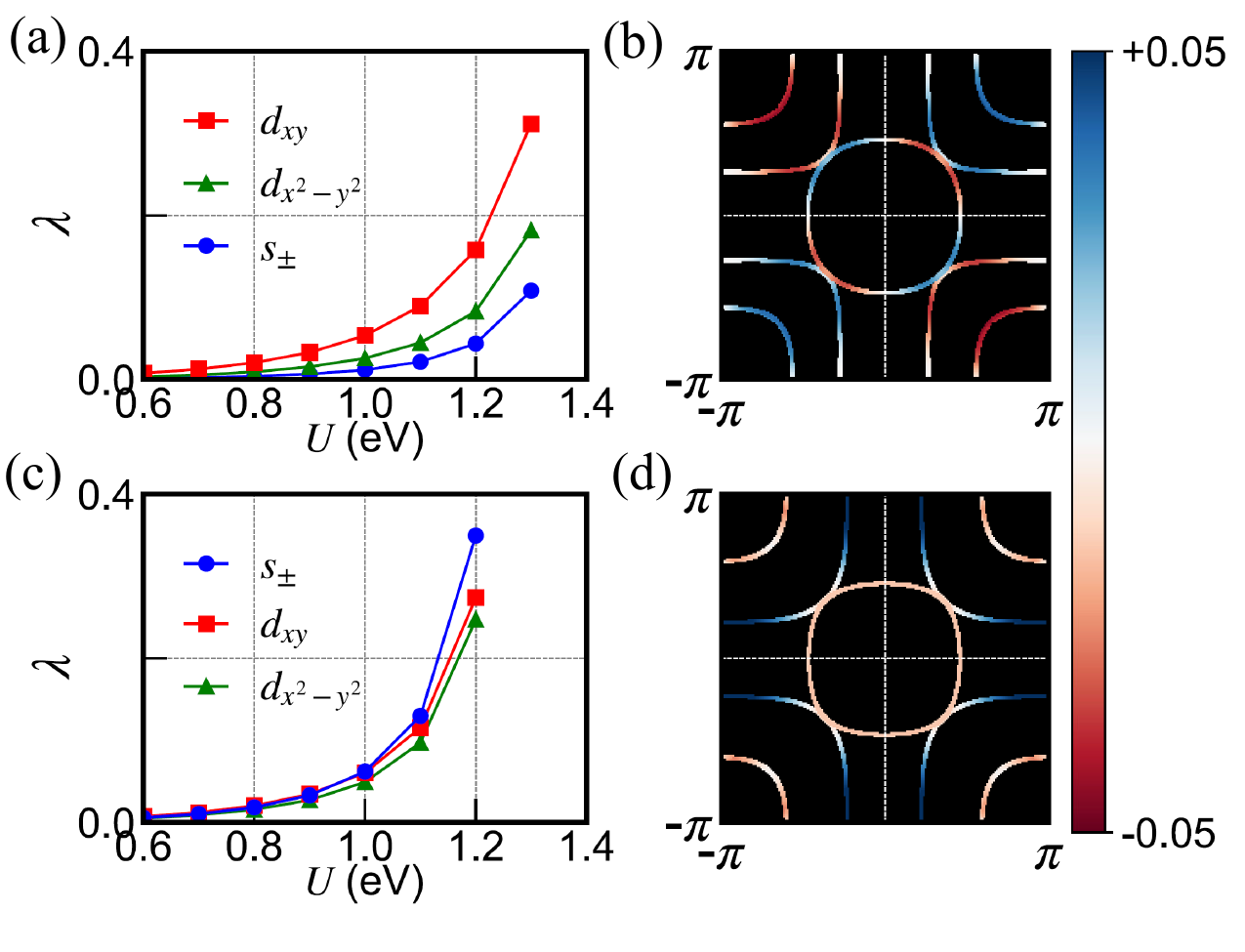}
\caption{Superconducting pairing symmetry according to the DFT + RPA approximation reported in Ref.~\cite{Liu:arxiv2023} for the La$_3$Ni$_2$O$_7$ nickelate material in bulk form, varying the crystal-field energy between the $d_{x^2-y^2}$ and $d_{3z^2-r^2}$ orbitals. This calculation indicates the dominance of $s^{\pm}$ or $d_{xy}$ depending on relatively small changes for the crystal-field energy in the tight-binding model, suggesting close competition between different pairing channels. Reprinted by permission from Springer Nature Customer Service Centre GmbH: Springer Nature, Nature Communications, ~\cite{Liu:arxiv2023}, Copyright (2024).}
\label{fig16}
\end{figure}

An important aspect of the physics of high-strain Ni-oxide thin films, in comparison to high-pressure bulk systems, is the role of the ratio $c/a$, where
$c$ is the out-of-plane lattice constant and $a$ is the in-plane lattice constant in the tetragonal geometry ($a = b$), as well as the role of the energy splitting (crystal field) between the two dominant Ni orbitals. Early work on the last subject in bulk materials indeed already pointed out~\cite{Liu:arxiv2023} that the dominant pairing symmetry (within the DFT + RPA approximation) depends sensitively on the crystal-field splitting between the two Ni orbitals, as shown in Fig.~\ref{fig16}. The crystal-field splitting effect for the $d$-wave to $s^{\pm}$-wave states, as well as Hund coupling $J_H$, were also discussed by constrained-path quantum Monte Carlo simulations~\cite{Xiong:arxiv25}. This aspect must also be carefully considered in theoretical studies of thin films, when deciding between the $d$- and $s^{\pm}$-pairing channels.

\begin{figure*}
\centering
\includegraphics[width=0.88\textwidth]{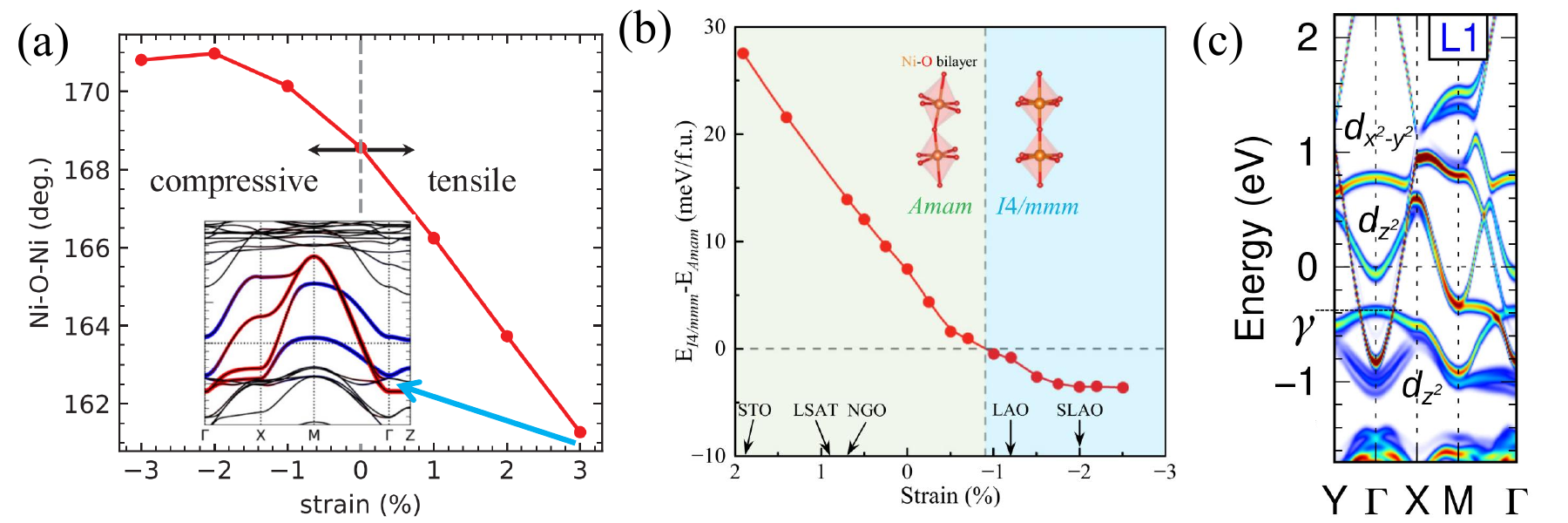}
\caption{(a) Changes in the angle of the Ni-O-Ni bond varying the in-plane lattice constants, while relaxing the out-of-plane constant via the introduction of strain.  Inset: band structure with tensile strain. These results are obtained on bulk format with strain effects. Reprinted with permission from~\cite{Zhao:prb25}, Copyright (2025) by the American Physical Society. (b) Influence of strain on the difference in energy between the $I4/mmm$ and $Amam$ phases, using {\it ab initio} DFT calculations. Five different tetragonal substrates are shown in the horizontal axis.  Reprinted with permission from  Ref.~\cite{Yi:PRB2025}, Copyright (2025) by the American Physical Society. (c) Band structure of interface bilayer for a (La$_3$Ni$_2$O$_7$)$_3$/(LSAO)$_3$ heterojunction. Reprinted with permission from  Ref.~\cite{Geisler:arxiv25}, Copyright (2025) by the American Physical Society.}
\label{fig17}
\end{figure*}

Among the early theoretical studies, the influence of strain on the electronic structure of bilayer nickelate bulk was investigated~\cite{Zhao:prb25}, with
the strain effects incorporated by modifying the in-plane lattice constants while allowing the out-of-plane lattice parameter $c$ to relax. It was found that compressive strain increases $c$, decreases $a$ and $b$, and shifts the $\gamma$ band below the Fermi level, whereas tensile strain produces the opposite effect, leading to the appearance of $\gamma$ pockets [see Fig.~\ref{fig17}(a)]. The suppression of the $\gamma$ band at the Fermi level tends to favor $d$-wave pairing over the competing $s^{\pm}$ state often obtained in DFT + RPA studies under high pressure. This is because the $s^{\pm}$ scenario typically relies on the presence of a $\gamma$ pocket at the Fermi level that can nest with the $\beta$ portion of the Fermi surface.

We now turn to the theoretical studies of bilayer thin films, focusing on the atomic and electronic band structures, beginning with {\it ab initio} approaches, such as DFT and its extension DFT + $U$, which incorporates on-site Hubbard interactions.

\begin{figure}
\centering
\includegraphics[width=0.48\textwidth]{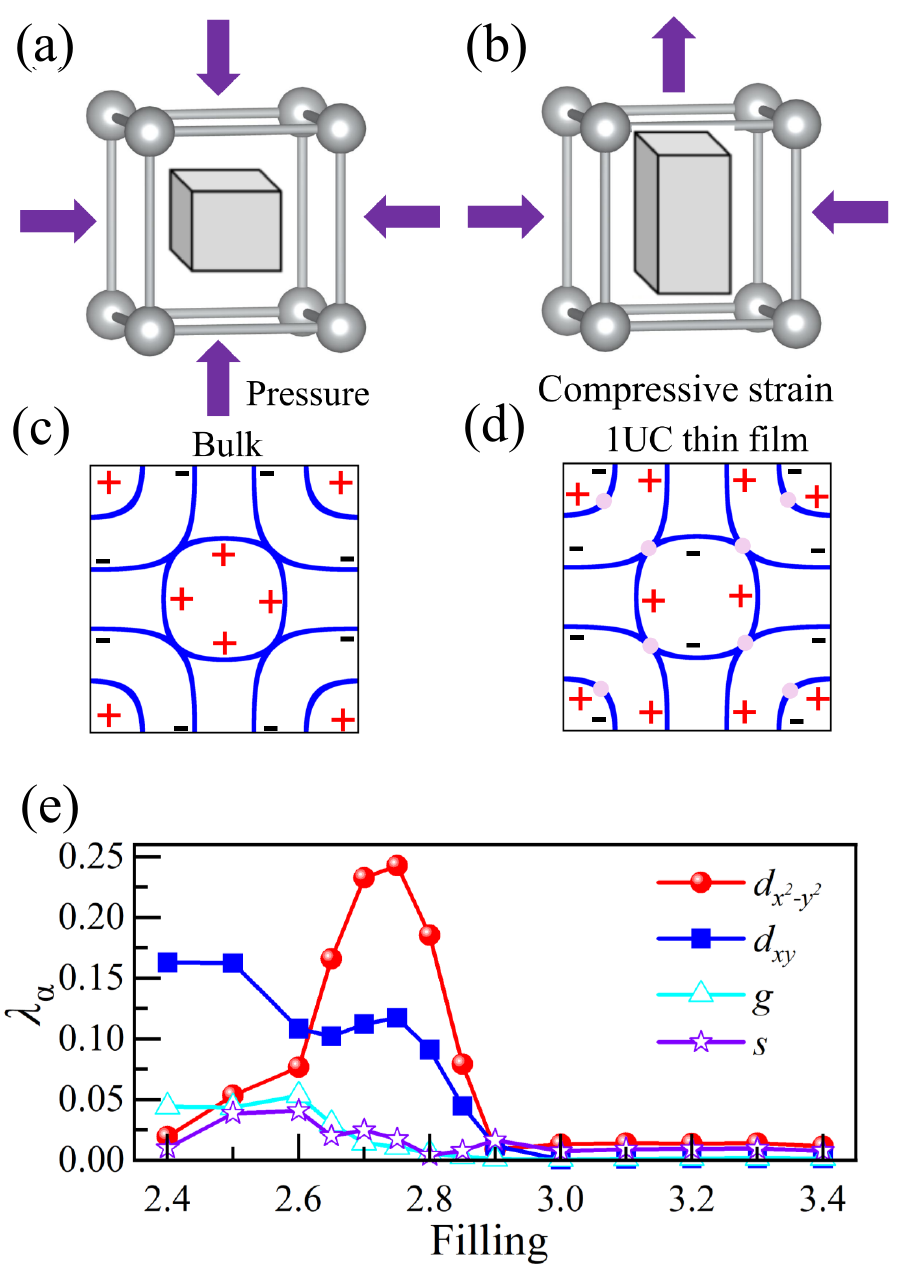}
\caption{(a-d) Sketch explaining the difference between hydrostatic pressure and compressive strain. (a-b) Schematic of lattice modifications under the high-pressure and in-plane compressive-strain thin-film conditions. (c-d) Sketches of the Fermi surfaces for the high-pressure ($s^{\pm}$) and compressive strain for hole-doping thin-films ($d_{x^2-y^2}$) with the signs of the superconducting order parameter. The light pink dots in (d) denote nodes. (e) The RPA calculated pairing strength $\lambda$ for different instability channels as a function of electron density fillings for the compressive-strain 1 UC case. Reprinted with permission from
Ref.~\cite{Zhang:arxiv2025-film}, Copyright (2026) by the American Physical Society.}
\label{fig18}
\end{figure}

\begin{itemize}

\item Ref.~\cite{Shi:cpl25} investigated the Fermi surface of bilayer nickelate films for varying thicknesses (from 0.5 to 3 UC) both with and without doping. Without doping, all cases have a $\gamma$ hole pocket at the M point, originating from the outer layers near the vacuum interface, while the inner bilayers do not contribute to this pocket. A similar observation was reported in Ref.~\cite{Hu:cp25}.

\item In Ref.~\cite{Li:arxiv20125-film}, DFT + $U$ calculations were performed for La$_3$Ni$_2$O$_7$ thin films grown on LSAO, including interfacial reconstruction, with a Hubbard $U$ value of 3.5 eV. The results support the hypothesis from an ARPES study that Sr interdiffusion at the interface provides intrinsic hole doping~\cite{Li:NSR25apres}.

\item In Ref.~\cite{Yi:PRB2025}, \textit{ab initio} calculations showed that compressive stain at $\sim 0.9 \%$ could stabilize $I4/mmm$, as displayed in Figure~\ref{fig18}(b). This work, considering hole doping presumably arising from Sr migration from LSAO, was also used to propose a unified framework connecting strain-driven and pressure-driven superconductivity.

\item In Ref.~\cite{Geisler:arxiv25}, bilayer nickelates on various substrates were investigated using DFT + $U$. The study revealed an unconventional occupation of the {\it antibonding} Ni 3$d_{3z^2-r^2}$ orbital at the $\Gamma$ point, giving rise to electron pockets, as displayed in Fig.\ref{fig17}(c). This offers a novel perspective because such observation has not been previously reported in studies of high-pressure bulk nickelates. This layer-resolved analysis near the interface also identifies a reconstructed interface composition, accompanied by a strong enhancement of spin fluctuations due to Fermi surface nesting associated with the antibonding Ni 3$d_{3z^2-r^2}$ band. Consequently, this work shifts the focus from the commonly discussed bonding hole pockets to the previously overlooked antibonding electron pockets, which may enhance the superconductivity as discussed recently in Ref.~\cite{Hua:arxiv2026}.

\item Ref.~\cite{Geisler:prb2025-film} explored the structural and electronic properties of La$_3$Ni$_2$O$_7$ on comprehensive LAO and tensile STO substrates.
Electron doping was found at the interface for the comprehensive LAO case, while no charge transfer was observed for the tensile STO substrate. Interestingly, tensile strain drives similar Fermi surface topology as discussed in high-pressure bulk La$_3$Ni$_2$O$_7$.

\end{itemize}

Based on the hopping parameters obtained from the band structure of DFT or DFT + $U$ calculations, typically supplemented by other methods to be able to address superconductivity, such as the RPA, several studies for Ni thin films were conducted. In most of these studies, only 0.5 UC containing a single bilayer was used to calculate the band structure for subsequent pairing model analyses.

\begin{itemize}

\item Ref.~\cite{Shao:prb25}, employing a single bilayer film with DFT + $U$ ($U=3.5$ eV) plus RPA, it was found that the bonding $d_{3z^2-r^2}$ band crosses the Fermi level with the appearance of a $\gamma$ pocket, leading to an $s^{\pm}$ pairing within RPA due to Fermi surface nesting reasons, like those of high-pressure bulk La$_3$Ni$_2$O$_7$. Upon hole doping, the $d_{xy}$ pairing state becomes dominant over $s^{\pm}$ state.

\item In another study~\cite{Cao:arxiv2025-film}, DFT combined with FRG calculations for single bilayer films indicate that, as the out-of-plane expansion increases, the $\gamma$ pocket characteristic of the high-pressure regime moves closer to the Fermi level. This shift enhances the density of states at the Fermi energy, leading to the conclusion that $s^{\pm}$ pairing remains robust in thin films, consistent with the bulk high-pressure behavior within certain theoretical approximations.

\item In Ref.~\cite{Shao:arxiv25}, using DFT + RPA calculations for La$_3$Ni$_2$O$_7$ single-bilayer films, pairing was found even without a $\gamma$ pocket, with the nesting between the $\alpha$ and $\beta$ sheets giving rise to an $s^{\pm}$ pairing state.

\end{itemize}

We now focus on approximations for thin films that can be considered to move beyond the weak coupling Hubbard $U$ regime, like RPA, or beyond the static DFT + $U$ method. For high-pressure bulk, strong coupling slave boson mean-field theory for a two-orbital $t-J$ model was used, also leading to $s^{\pm}$ pairing~\cite{Ji:prb25-strong}. This provides a smooth connection between weak and strong coupling, as well as between high-pressure and high-strain samples, at least within the approximations used

\begin{itemize}

\item In Ref.~\cite{Yue:NSR25}, a combination of \textit{ab initio} DFT calculations and constrained random phase approximation (cRPA) was employed within a multiorbital Hubbard model for thin films. In addition, cluster dynamical mean-field theory (CDMFT) was used to capture correlation effects beyond the weak-coupling level. By comparison with ARPES experiments, the authors identified the appropriate electronic filling for the theoretical analysis. Density functional theory plus Hubbard $U$ (DFT + $U$) calculations were also performed. Overall, a pronounced $s^{\pm}$ pairing instability was found, driven by Fermi surface nesting with dominant $d_{3z^2-r^2}$ orbital character, in close analogy to the bulk system under high pressure.

\item In another work, using self-consistent mean-field calculations based on a model tailored for La$_{2.85}$Pr$_{0.15}$Ni$_2$O$_7$, it was also reported $s^{\pm}$ interlayer pairing as the dominant superconducting channel~\cite{Huang:arxiv2025-pairing}. Using a renormalized mean-field approach applied to a two-orbital bilayer $t–J$ model, $s^{\pm}$ pairing was also observed~\cite{Qiu:arxiv25}. The results are consistent with ARPES measurements~\cite{Shen:arxiv2025}, which indicate the absence of nodes in the $\beta$ and $\gamma$ pockets.

\item In Ref.~\cite{Bleys:arxiv25}, using a fully charge self-consistent DFT plus embedded dynamical mean-field theory (eDMFT) approach to study La$_3$Ni$_2$O$_7$ under several compressive strains, the $\gamma$ pocket was found to emerge and cross the Fermi level, exhibiting flat band features when dynamical correlations are included.

\item In Ref.~\cite{Zhong:arxiv25}, magnetic and charge distribution properties of models for La$_3$Ni$_2$O$_7$ thin films, involving up to 11 bands within a Hubbard model, were investigated using determinant quantum Monte Carlo techniques.

\end{itemize}

A generic comment affecting the efforts where hoppings are obtained in 0.5 UC film models: note that the case of 0.5 UC corresponds to a single bilayer. The more realistic case of 1 UC has two bilayers (namely, four NiO$_2$ planes), one shifted by 1/2 in-plane lattice spacing with respect to the other.  Specifically, the DFT unit cell used consists of vacuum–layer1–layer2–vacuum, with the in-plane lattice constants ($a$ and $b$) fixed to those of LSAO and the out-of-plane constant ($c$) adjusted automatically. As a result, modeling thin films with only 0.5 UC in the above-described setup does not realistically capture the asymmetry present in actual films, which are sandwiched between a substrate and a vacuum. For thicker slabs beyond 0.5 UC, where an asymmetry between the NiO$_2$ planes can be established, realistic simulations become possible.

In Ref.~\cite{Zhang:arxiv2025-film}, our group investigated a 1 UC La$_3$Ni$_2$O$_7$/LSAO thin film using DFT + RPA methods. As expected, the interlayer hopping amplitude is reduced (specifically by approximately $15 \%$) compared to the bulk, due to the expansion of the $c$-axis lattice parameter. Our results also indicate that, in the undoped limit, $\gamma$ pockets are absent, leading to a lack of significant Fermi surface nesting and, consequently, no pairing instability within RPA. Upon increasing hole doping, however, pairing tendencies emerge predominantly in the $d$-wave channel. Figures~\ref{fig18}(a-d) provide a qualitative illustration of the key differences between the high-pressure bulk and the compressively strained thin-film cases. Remarkably, depending on the physical conditions, i.e., bulk crystals under high pressure vs. thin films under high compressive strain, the same material may favor either an $s^{\pm}$ or a $d$-wave pairing tendency, at least within our approximations.

Figure~\ref{fig18}(e) shows the RPA pairing strength $\lambda$ as a function of electron density $n$ (with $n=3$ corresponding to the undoped case) for various pairing channels~\cite{Zhang:arxiv2025-film}. At $n=3$, no pairing instability is observed. Upon hole doping to $n\sim 2.8$ (approximately $6.66 \%$ hole doping), the $d_{x^2-y^2}$ channel dominates, whereas at $n\sim2.5$ the $d_{xy}$ channel becomes dominant. It would be interesting to examine whether this weak-coupling analysis, which is sensitive to Fermi surface details, persists in non-perturbative calculations beyond weak coupling. In this study, the RPA magnetic order is dominated by a wavevector $\mathbf{q}\sim(\pi/3,\pi/3)$, indicating diagonal stripe order as the magnetic precursor to the pairing instability.

A related theoretical study, using DFT and FRG techniques, based on 3 UCs grown on a (Sr, La)AlO$_3$ substrate (note, not on LSAO$_4$), shows that the interfacial bilayer is hole-doped and develops $s^{\pm}$ pairing, reinforced by competing spin-density wave fluctuations~\cite{Le:arxiv20125-film}. The maximum $T_c$ is obtained at the Lifshitz transition when the $\gamma$ pockets appear at the Fermi surface~\cite{Le:arxiv20125-film}. Furthermore, an interesting theoretical proposal establishes the connection between the critical temperature of superconductivity in bulk La$_3$Ni$_2$O$_7$ and its film under different conditions~\cite{Chen:arxiv2026film}. In addition, they also proposed that electronic doping, or further enhancement of the compressive strain in the film, can enhance $T_c$.

\begin{figure}
\centering
\includegraphics[width=0.48\textwidth]{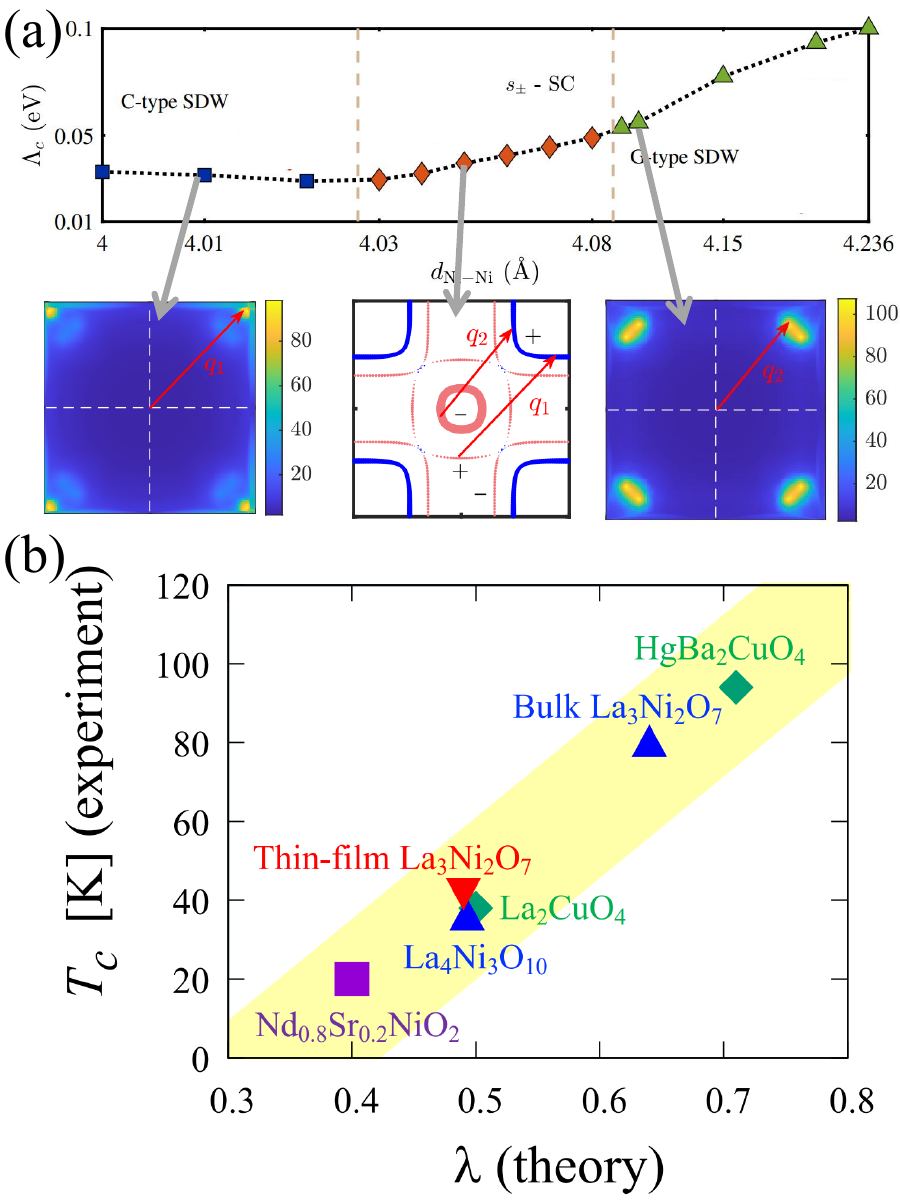}
\caption{(a) The calculated transition temperature of the ordered states vs. the nearest-neighbour $d_{Ni-Ni}$ distance along the $z$-axis for La$_3$Ni$_2$O$_7$ thin films grown on the LaAlO$_3$ substrate, with the sketch of three states. Left: C-type SDW in momentum space at $d_{Ni-Ni} = 4.01$ \AA; middle: superconducting order parameter with $s^{\pm}$ symmetry at $d_{Ni-Ni} = 4.05$ \AA; right: G-type SDW in momentum space at $d_{Ni-Ni} = 4.10$ \AA. Reprinted from Ref~\cite{Cao:arxiv2026-lao}, CC BY 4.0. (b) The superconducting transition temperature $T_c$ from experiments is plotted as a function of the theoretical RPA pairing strength $\lambda$, using the FLEX approximation. Reproduced from\cite{Ushio:arxiv25}. Reprinted from Ref.~\cite{Ushio:arxiv25}, CC BY 4.0.}
\label{fig19}
\end{figure}

Ref.~\cite{Cao:arxiv2026-lao} studies La$_3$Ni$_2$O$_7$/LaAlO$_3$ using FRG, and an $s^{\pm}$ pairing state was found. In addition, they also proposed that the interlayer Ni-Ni distance could transform the ground state from C-type SDW (layer-even SDW), first to an $s^{\pm}$-wave pairing channel, and then to G-type SDW (layer-odd SDW), with increasing interlayer distance, as shown in Fig.~\ref{fig19}(a).

In Ref.~\cite{Ushio:arxiv25}, a fluctuation exchange approximation (FLEX) was applied to bulk systems with strain, yielding a robust $s^{\pm}$ pairing symmetry found to be largely insensitive to details of the band structure. FLEX incorporates frequency-dependent interactions and self-energy effects, as compared to the Fermi surface-restricted RPA. Within this framework, finite Hubbard $U$ and Hund’s coupling $J$ are included.

Figure~\ref{fig19}(b) presents a comparison of the pairing strength between nickelates and cuprates. In this framework, La$_3$Ni$_2$O$_7$ nickelate thin films are found to be close to the cuprate 214 family, at least at the level of the FLEX approximation that was used in this analysis. These results suggest that nickelates likely reside in the intermediate- to strong-coupling regime with respect to the Hubbard interaction $U$, by analogy with the cuprate 214 compounds, for which this parameter range is relatively well established. However, this conclusion relies on the assumption that superconductivity is driven by magnetic fluctuations and that the RPA-based pairing strengths can be meaningfully compared across different materials. The latter point is nontrivial, given that different values of $U$ are typically adopted for each system. Therefore, these results should be interpreted as qualitative.

\begin{figure}
\centering
\includegraphics[width=0.48\textwidth]{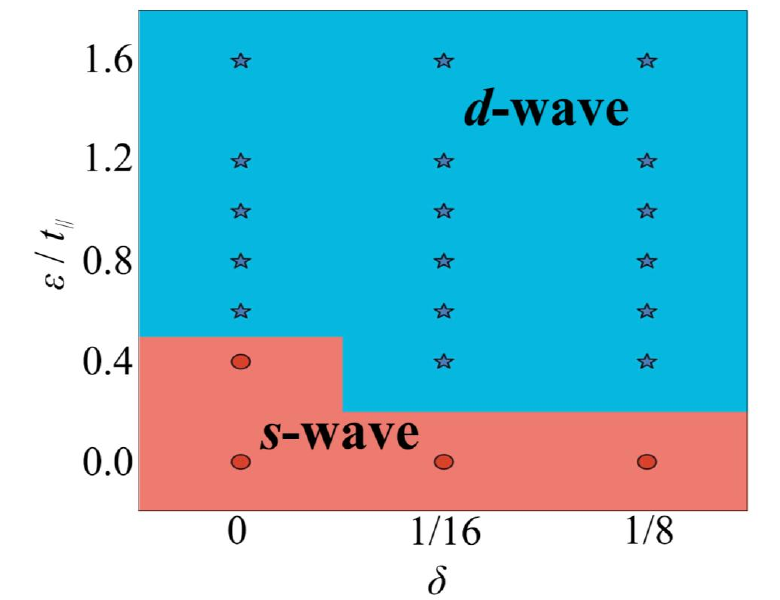}
\caption{Phase diagram of the single-bilayer film on LSAO, under the influence of an electric field $\epsilon$ perpendicular to the film. $\delta$ is the electron doping level of the $d_{x^2-y^2}$ orbital. Reprinted with permission from Springer Nature Customer Service Centre GmbH: Springer Nature, Nature Communications,~\cite{Shao:nc26}, Copyright (2026).}
\label{fig20}
\end{figure}

Finally, we discuss a novel theoretical proposal involving an electric field applied perpendicular to the films~\cite{Shao:nc26}, which may substantially enhance the superconducting transition temperature $T_c$. As noted above, in a vacuum-bilayer-vacuum geometry, the two layers in a single bilayer are equivalent by symmetry. However, this symmetry can be broken by applying a perpendicular electric field, which induces charge transfer between the two Ni layers as well as redistribution among orbitals. The net effect, as reported in Ref.~\cite{Shao:nc26}, is a suppression of the $s$-wave pairing associated with the bonding band and a concomitant enhancement of $d$-wave pairing (see Fig.~\ref{fig20}). These results are obtained based on calculations using the $t–J$ model, employing both slave-boson mean-field theory and density matrix renormalization group simulations on a $2\times2\times64$ cluster.

In summary, the theoretical description of thin films is more challenging than that of bulk materials. The reduced dimensionality and the film geometry render the Ni planes of the thin films inequivalent, adding complexity to their modeling. Most theoretical studies of Ni-oxide thin films continue to favor $s^{\pm}$ pairing. However, some results support a $d$-wave state for thin films, even in cases where the same many-body approaches were applied before to high-pressure bulk systems yielding $s^{\pm}$ pairing.

\section{Superconductivity in hybrid stacking RP nickelates}
\label{sec_6}
\vspace{6pt}

While experimental efforts continue to improve the sample quality, transition temperature $T_c$, and superconducting volume fraction for both bulk and film RP nickel oxides, an interesting alternating single-layer
and trilayer stacking structure (La$_2$NiO$_4$)/(La$_4$Ni$_3$O$_{10}$) was synthesized experimentally in bulk format, with the same chemical formula La$_3$Ni$_2$O$_7$~\cite{Chen:jacs,Puphal:arxiv12,Wang:ic,Abadi:arxiv24}, by several groups independently. This is called the 1313 nickelate. At the same time, another hybrid RP nickelate with alternating single-layer and bilayer stacking (La$_2$NiO$_4$)/(La$_3$Ni$_2$O$_{7}$) was also prepared in bulk format~\cite{Li:prm24}, namely 1212 nickelate. Similar to the RP nickel oxides, superconductivity is not observed in those hybrid stacking RP nickelates without pressure.

\begin{figure*}
\centering
\includegraphics[width=0.88\textwidth]{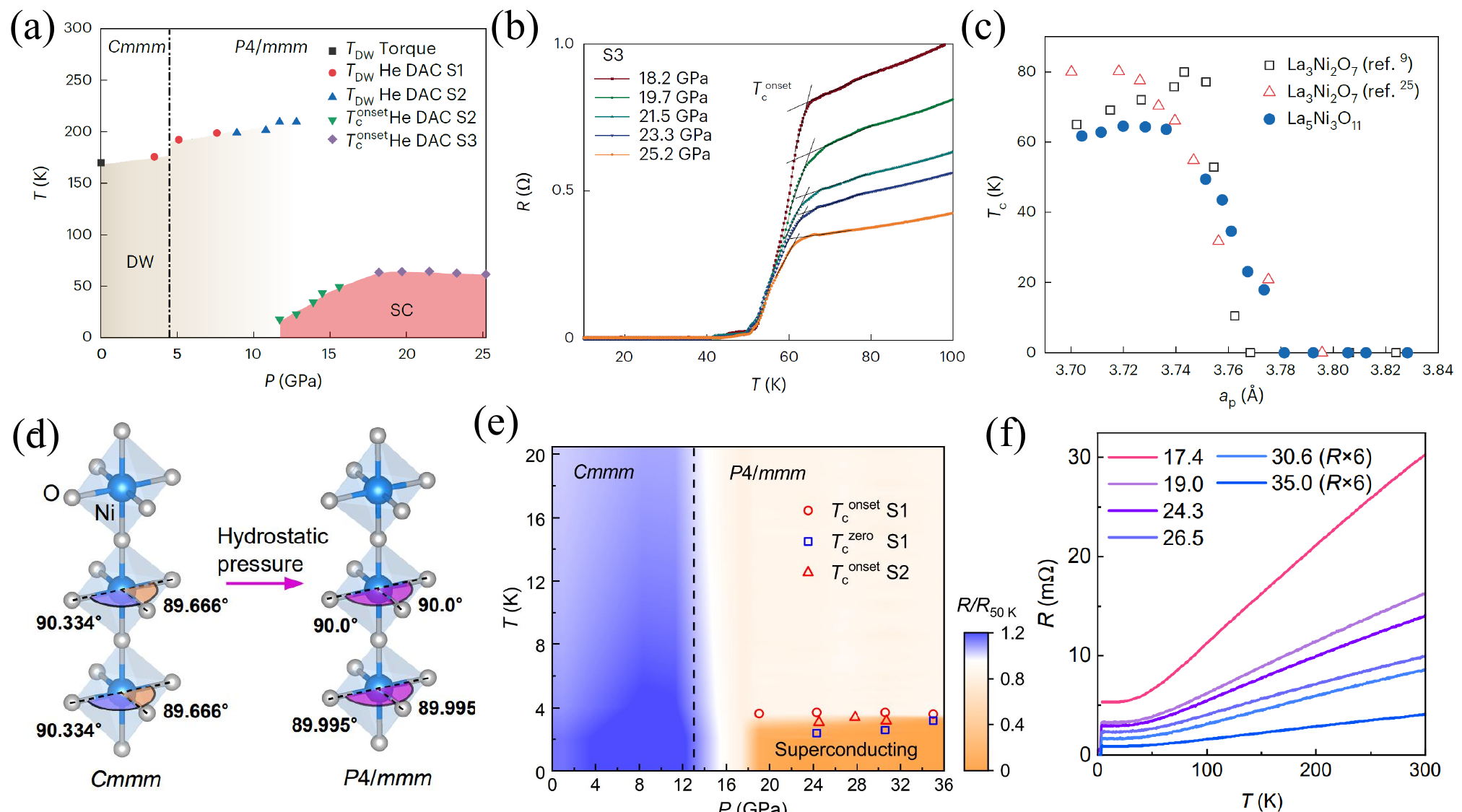}
\caption{(a) Pressure-dependent phase diagram showing the crystal structures transitions, the density-wave (denoted DW) transitions, and superconductivity transitions for 1212 nickelate single crystals. (b) Resistance vs. temperature $R(T)$ curves of 1212 nickelates (Sample S3) single crystal for different pressures. (c) The relationship between $T_c$ and the average in-plane lattice ($a_p$) for 1212 nickelates. (a-c) are reprinted by permission from Springer Nature Customer Service Centre GmbH: Springer Nature, Nature Physics, ~\cite{Shi:arxiv25}, Copyright (2025). (d)  Schematic of the trilayer crystal structure portion of the 1313 phase, before and after the structural transition. (e) Pressure-dependent phase diagram for the 1313 nickelate. (f) Resistance vs temperature $T$ curves of 1313 nickelates for different pressures. (d-f) are reprinted from ~\cite{Huang:arxiv25-1313}, CC BY 4.0.}
\label{fig21}
\end{figure*}

Surprisingly, superconductivity has recently been observed in the 1212 nickelate under pressure~\cite{Shi:arxiv25}. This 1212 nickelate shows an orthorhombic $Cmcm$ structure with slightly tilted NiO$_6$ octahedra at ambient conditions. As shown in Fig.~\ref{fig21}(a), it also exhibits a density-wave transition at the ambient and low-pressure regions, similar to bilayer La$_3$Ni$_2$O$_7$. With increasing pressure, around 5 GPa, the $Cmcm$ phase transfers to the high-symmetry $P4/mmm$ phase with untilted NiO$_6$ octahedra, followed by the emergence of superconductivity at around 12 GPa to 25 GPa (the maximum pressure they studied). The maximum onsite $T_c \sim 64$ K was found at 18.2 GPa, as displayed in Fig.~\ref{fig21}(b), while zero resistance was obtained at $\sim 40$~K. The optimal superconductivity with large superconducting volume fraction was observed at approximately 21 GPa with $T_c = 54$~K. Interestingly, Fig.~\ref{fig21}(c) shows $T_c$ as a function of the average in-plane lattice constant $a_p$ for the 1212 nickelate. It appears to be a ``critical $a_p$'' close to 3.8~\AA, below which superconductivity emerges, which is quite similar to that for high-pressure bulk La$_3$Ni$_2$O$_7$.

More recently, the 1313 nickelate was also reported to superconduct under pressure, albeit at a significantly reduced temperature of 3.6 K~\cite{Huang:arxiv25-1313}.  Pressure induces the structural transition from a slightly distorted orthorhombic $Cmcm$ structure to the tetragonal $P4/mmm$ phase with Ni-O-Ni $180^\circ$ around 13 GPa, as shown in Fig.~\ref{fig21}(d). Similar to 1212 nickelates, the emergence of superconductivity under presure does not precisely coincide with the structural phase transition [see Fig.~\ref{fig21}(d)]. A sharp drop in resistance appears with onset $T_c$ $\sim 3.6$ K at 19 GPa [see Fig.~\ref{fig21}(e)], and a clear zero-resistance state below 2.3 K was observed around 24.3 GPa.

\begin{figure*}
\centering
\includegraphics[width=0.88\textwidth]{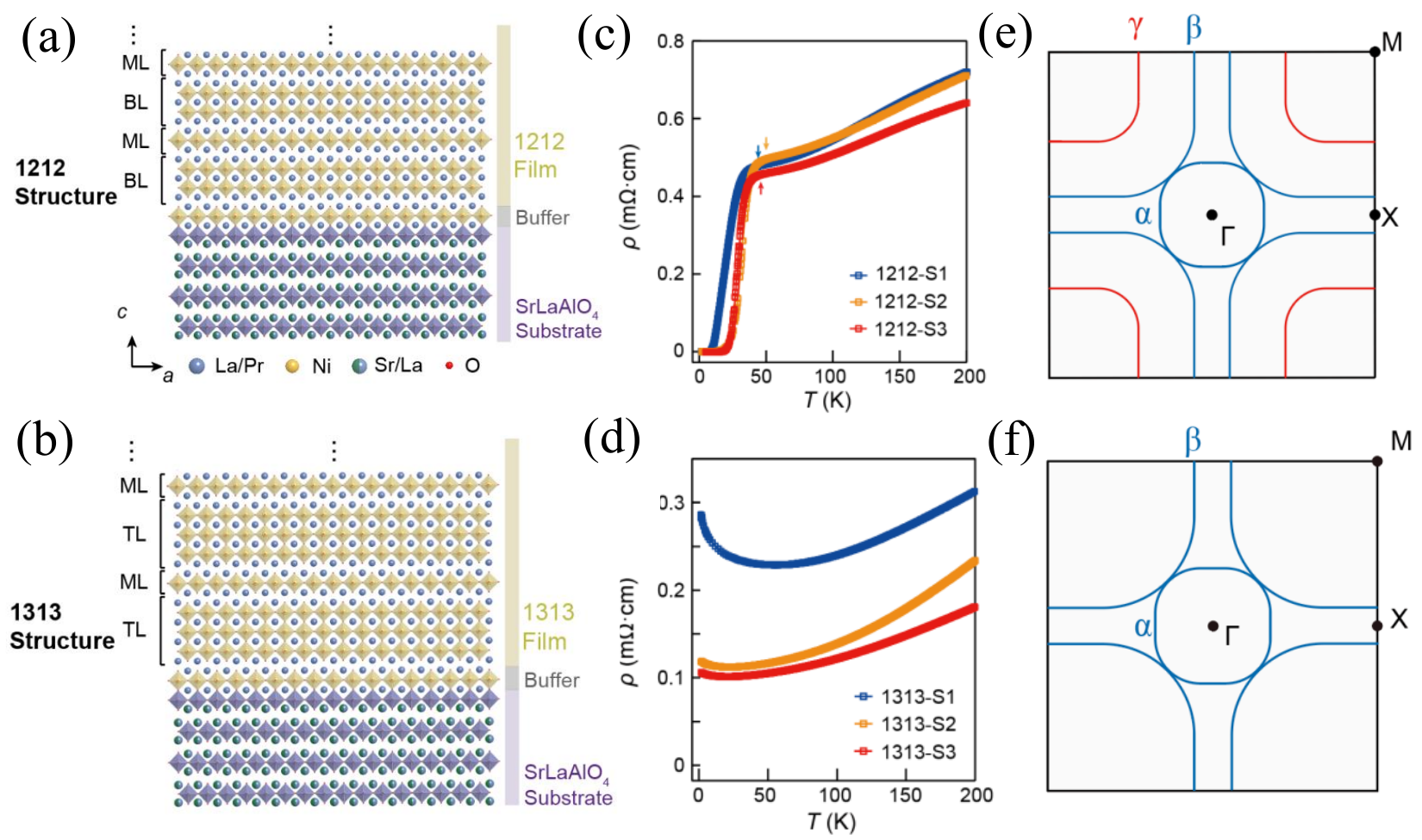}
\caption{Structural schematic of the (a) 1212 and (b) 1313 hybrid films on the LSAO substrate. ML and BL
represent monolayer and bilayer, respectively. Temperature-dependent resistivity curves for (c) 1212 and (d) 1313 nickelate films
for different samples. Fermi surface schematic for (e) 1212 and (f) 1313 nickelate films. Reprinted by permission from Springer Nature Customer Service Centre GmbH: Springer Nature, Nature, ~\cite{Nie:arxiv2025-apres}, Copyright (2019).}
\label{fig22}
\end{figure*}

While studying the hybrid stacking RP nickelates in bulk form, 1212 and 1313 nickelate thin films were also prepared, growing on a LSAO substrate~\cite{Nie:arxiv2025-apres}, as displayed in Figs.~\ref{fig22}(a-b). At ambient pressure, superconductivity was
found in the 1212 nickelate film with onset $T_c$ up to 50 K, as shown in Fig.~\ref{fig22}(c). However, the 1313 nickelate film does not show superconductivity at ambient conditions (see Fig.~\ref{fig22}(d)). The sketches of the ARPES results (see Fig.~\ref{fig22}(e)) for the bilayer part of the 1212 nickelate film show that the $\gamma$ pocket crosses the Fermi level, leading to similar Fermi surface topology as in bilayer nickelate films. In contrast, Fig.~\ref{fig22}(f) presents the Fermi surface of the trilayer part of 1313 nickelate film, where the $\gamma$ pocket is absent.

To finalize this section, we summarize below other recent theoretical progress for the hybrid stacking RP nickelates.

\begin{itemize}

\item 1212 nickelates: Ref.~\cite{Zhang:1212} found the ``charge transfer'' between the single-layer and bilayer sublattices, leading to hole doping in the single-layer part. In addition, Ref.~\cite{LaBollita:1212} found that 1212 nickelates show layer-selective physics with the  Mott instability in single layers, while the bilayers contribute to its low-energy physics. RPA calculations provide a leading $d_{x^2-y^2}$ channel arising mainly from the single-layer blocks at smaller $U$ and a leading $s^{\pm}$ instability from the bilayer part at larger $U$ (similar to high-pressure bulk La$_3$Ni$_2$O$_7$)~\cite{Zhang:1212}. Ref.~\cite{Zhang:arXiv2506-1212} also reported $s^{\pm}$ pairing from the bilayer subsystem and argued the single-layer subsystem mainly serves as a bridge facilitating the inter-bilayer phase coherence through the interlayer Josephson coupling.

\item 1313 nickelates: Ref.~\cite{Zhang:1313} reported that the $Cmcm$ and $P4/mmm$ phases exhibit nearly identical electronic structures, with the $\gamma$ pocket absent in the trilayer region under both ambient and high pressures. Their results also suggest that the superconducting transition temperature ($T_c$) of the 1313 structure is unlikely to reach values high enough to match the experimentally observed $T_c$ in La$_3$Ni$_2$O$_7$~\cite{Zhang:1313}. Ref.~\cite{LaBollita:1313} reported that the Mott gap in the single-layer block closes due to orbital-selective physics, leading to transfer of charge into the trilayer block under pressure. Similar conclusions of layer selectivity and single-layer Mott critical behavior were also obtained in Ref.~\cite{Lechermann:PRM2024}. In addition, a characteristic Fermi-arc feature was found in the high-pressure 1313 phase, arising from the strongly correlated interplay of the multi-orbital and multilayer electronic structure, along with a concomitant low-energy $d_{3z^2-r^2}$ flat-band feature~\cite{Lechermann:PRM2024}.

\item 2323 nickelates: Ref.~\cite{Zhang:2323} predicted a leading $s^{\pm}$ pairing state for bilayer-trilayer stacking nickelates, namely the 2323 structure,  under pressure with a similar or higher superconducting transition temperature $T_c$, as obtained in the bilayer La$_3$Ni$_2$O$_7$. While finishing the present review, 2323 stacking nickelate thin films have been prepared and are confirmed to be superconducting at ambient conditions with an onset $T_c$ around 46 K~\cite{Nie:arxiv2025-apres}.

\end{itemize}

In summary, the new direction of studies involving hybrid stacking RP nickelates extends the area of nickel oxides,
providing a new platform for research in thin films, to grow samples layer-by-layer. Based on present efforts, the strong superconducting pairing in 1212 nickelates seems to originate from bilayer blocks, with the appearance of the $\gamma$ pocket. In addition, the trilayer is the conducting layer in the 1313 structure, where the $\gamma$ pockets from trilayer blocks are also absent. More studies are still needed to clarify these interesting points.

\section{Summary and outlook}
\label{sec_7}
\vspace{6pt}

The discovery of superconductivity with $T_c \sim 80$ K in the bilayer RP nickelate La$_3$Ni$_2$O$_7$ bulk was initially reported in 2023. This exciting experiment opened a remarkable new platform for the study of the origin of unconventional superconductivity, unveiling several challenging results that theory needs to explain. To date, among all RP nickelate family materials and related hybrid stacking variations, four bulk systems (bilayer, trilayer, 1212, and 1313) have been found to be superconducting under high pressure, while three film systems (bilayer, 1212, and 2323) have confirmed superconductivity at ambient pressure as well.

In Sec.3, we reviewed the present status of bulk bilayer nickelates, in a reduced format because the field is too vast and
the focus of this review is the thin films.
Within bulk bilayers, several key issues remain to be addressed: (1) Fermi surface topology under pressure: although nearly all theoretical studies predict the presence of $\gamma$ pockets in RP bilayer nickelates under pressure, experimental confirmation is still lacking due to the challenges of performing high-pressure ARPES measurements. (2) Magnetic structure: although some RIXS experiments provide evidence for the double stripe spin state in the $Amam$ phase at ambient pressure, the magnetic structure has not yet been fully established and remains under debate. Because of sample size and quality, direct inelastic neutron scattering experiments in single crystals are still needed to identify the intrinsic magnetic structure. As discussed Sec.3.4, the structural distortion itself favors the double spin stripe. Under high pressure, the structural distortion disappears, and the high-pressure magnetic structure may be different from that at ambient conditions. Thus, high-pressure experiments are needed to understand the relationship between the superconducting phase and the magnetic state.

In Sec.4 and  Sec.5, we reviewed in detail recent progress on bilayer ultra-thin films. The experimental results for the Fermi surface of the nickelate films are still inconclusive on (1) whether the $\gamma$ pockets appear at the Fermi level, and (2) whether the superconducting state does or does not have nodes. This may relate to scenarios for accidental carrier doping in thin films of La$_3$Ni$_2$O$_7$ due to Sr migration from the sample, oxygen non-stoichiometry, intrinsic interfacial charge transfer, or other possibilities. For the theoretical part, the $s^{\pm}$ leading solution is supported by many studies on the 0.5 UC structural model. However, a slab model restricted to 0.5 UC does not capture the symmetry breaking of the interface. Interestingly, our recently published work based on the 1 UC model that includes symmetry breaking of the two Ni layers in the bilayer lattice provides a $d$-wave solution instead.

In Sec.6, we reviewed recent progress on hybrid stacking RP nickelates. This direction is still waiting to be explored, and only a few groups are working on it. Present efforts suggest that the main physics of the 1212 and 1313 nickelates are related to their
bilayer and trilayer blocks. Thus, further study on those families of thin film systems may help understand the bilayer and trilayer RP nickelate in bulk as well.

In addition, thin films of the high-order nickelates La$_{m+1}$Ni$_m$O$_{3m+1}$ ($m = 4$ and $m = 5$)~\cite{Li:apl20} and Nd$_{m+1}$Ni$_m$O$_{3m+1}$ ($m = 4$ and $m = 5$)~\cite{Sun:prb21} have been stabilized in experiments using reactive molecular-beam epitaxy. Considering the present progress by providing additional pressure on bilayers, this could be a feasible way to explore superconductivity in high-order nickelate systems.

\section{Acknowledgments}
We are grateful for fruitful discussions with Adriana Moreo, Satoshi Okamoto, Harold Y. Hwang, Michael S. Norman, Antia S. Botana, Gang Su, Fan Yang, Meng Wang, Hai Hu Wen, Yijun Yu, Qianghua Wang, Huiqiu Yuan, Yuefeng Nie, Jinguang Cheng, Peter J. Hirschfeld, Benjamin Geisler, Dao-Xin Yao, Zhihai Zhu, Xingjiang Zhou, Zhuoyu Chen, and Qi-Kun Xue.

The authors are supported by the U.S. Department of Energy, Office of Science, Basic Energy Sciences, Materials Sciences and Engineering Division.  This manuscript has been authored by UT-Battelle, LLC, under contract DE-AC05-00OR22725 with the US Department of Energy (DOE). The US government retains, and the publisher, by accepting the article for publication, acknowledges that the US government retains a nonexclusive, paid-up, irrevocable, worldwide license to publish or reproduce the published form of this manuscript, or allow others to do so, for US government purposes. DOE will provide public access to these results of federally sponsored research in accordance with the DOE Public Access Plan (https://www.energy.gov/DOE Public Access Plan).

\section{Data Availability}
This review article contains no original data, and the cited references should be consulted for the original data.

\end{document}